\documentclass[]{elsarticle}
\usepackage{amsmath}
\usepackage{graphicx}
\usepackage{color}
\newcommand{\be}{\begin{equation}}

\newcommand{\ee}{\end{equation}}
\newcommand{\bea}{\begin{eqnarray}}
\newcommand{\eea}{\end{eqnarray}}

\newcommand{\de}{\partial}

\newcommand{\gapless}{\theta(\delta\mu-\Delta)}

\newcommand{\threeier}[1]{\int \frac{d^3#1}{(2\pi)^3}}
\newcommand{\intspace}[1]{\int d^4 {#1}\,}
\newcommand{\ha}{\frac{1}{2}}

\newcommand{\vk}{{{\bf k}}}

\newcommand{\vp}{{{\bf p}}}

\newcommand{\tila}{\tilde{A}}
\newcommand{\tileta}{\tilde{\eta}}
\newcommand{\tiletac}{\tilde{\eta^*}}
\newcommand{\mdm}{(\delta\mu\rightarrow-\delta\mu)}
\newcommand{\xip}{\xi({\bf p})}
\newcommand{\dmu}{\delta\mu}
\newcommand{\dm}{\delta\mu}
\newcommand{\tr}{{\rm{Tr}}}

\newcommand{\Pd}{\Psi^\dagger}

\newcommand{\bmu}{\bar{\mu}}
\newcommand{\bdmu}{\delta\bar{\mu}}
\newcommand{\bfA}{{\bf{A}}}
\newcommand{\magp}{{\rm{p}}}

\begin{document}
\title{Collective modes in asymmetric ultracold Fermi systems}
\date{\today}
\author[MIT,ITEP]{Elena~Gubankova}
\ead{elena1@mit.edu}
\author[CTP,ECM]{Massimo~Mannarelli}
\ead{massimo@ieec.uab.es}
\author[CTP,LANL]{Rishi~Sharma~\corref{cor1}}
\ead{rishi@lanl.gov}
\date{\today}
\address[MIT]{Massachusetts Institute of Technology, Cambridge, MA 02139, USA}
\address[ITEP]{Institute of Theoretical and Experimental Physics, B.
Cheremushkinskaya 25, RU-117 218 Moscow, Russia}
\address[CTP]{Center for Theoretical Physics, 
Massachusetts Institute of Technology, USA}
\address[ECM]{Departament d'Estructura i Constituents de la Mat\`eria and
Institut de Ci\`encies del Cosmos (ICCUB), Universitat de Barcelona,
Mart\'i i Franqu\`es 1, 08028 Barcelona, Spain}
\address[LANL]{Theoretical Division, Los Alamos National Laboratories, 
Los Alamos, NM, 87545, USA}
\cortext[cor1]{Corresponding author. Phone:+1(505)606-2131. Fax:+1(505)667-1931.}

\begin{abstract}
 We derive the long-wavelength effective action for the collective modes in systems of
fermions interacting via a short-range $s$-wave attraction,
featuring unequal chemical potentials for the two fermionic species (asymmetric
systems). As a consequence of the attractive interaction, fermions form a
condensate that spontaneously breaks the $U(1)$ symmetry associated with total
number conservation. Therefore at sufficiently small temperatures and
asymmetries, the system  is a superfluid.
We reproduce previous results for the stability conditions of the system as a function
of the four-fermion coupling and asymmetry. We obtain these results analyzing the coefficients
of the low energy effective Lagrangian of the modes describing fluctuations in the magnitude
(Higgs mode) and in the phase (Nambu-Goldstone, or Anderson-Bogoliubov, mode) of the difermion condensate. 
We find that for certain values of parameters, the  mass of the Higgs mode decreases with
increasing mismatch between the chemical potentials of the two populations, if
we keep the scattering length and the gap parameter constant. Furthermore, we
find that the energy cost for creating a position dependent fluctuation of the
condensate is constant in the gapped region and increases in the gapless region. These two features may lead to experimentally
detectable effects. As an example, we argue that if the superfluid is put in
rotation, the square of the radius of the outer core of a vortex should 
sharply increase on increasing the asymmetry, when we
pass through the  relevant region in the gapless superfluid phase. Finally,
by gauging the global $U(1)$ symmetry, we relate the coefficients of the
effective Lagrangian of the Nambu-Goldstone mode with the screening masses of the
gauge field.
\end{abstract}
\maketitle

\section{Introduction}

 Experiments  with trapped cold atomic gases have driven a renewed interest in
fermionic pairing \cite{ketterle,Partridge}. In particular, much effort has been
devoted to understanding the superfluid phases of imbalanced fermionic gases,
featuring unequal number of particles of the distinct fermionic species that
pair~\cite{Clogston:1962,Sarma,LOFF,Muther:2002mc,Bedaque:2003hi,
Liu:2002gi,Gubankova:2003uj,Forbes:2004cr,Carlson:2005kg,Castorina:2005kg,
Pao,Pao:2006,Son:2005qx,Sheehy:2006qc,Yang:2005,trap,Chen,Bulgac:2006gh,Yang:2006ez,
Gubankova:2006gj,Mannarelli:2006hr,Carlson:2007}.

 The system consists of fermions of two different species, $\psi_1$ and
$\psi_2$, which correspond to two hyperfine states of a fermionic atom like
$^6{\rm Li}$. These fermions have opposite spin and the interaction between them
can be tuned by employing a Feshbach resonance \cite{Feshbach}. The strength of
the interaction is given in terms of the $s$-wave scattering length between the
two species.

 For zero imbalance, the system properties are qualitatively well understood
using mean field theory~\cite{BEC}. In weak coupling the system lives in a
weakly coupled BCS state and crosses over to a strongly coupled BEC state
through the resonance region. While the extreme BCS and BEC regimes are also in
good quantitative control in mean field theory, close to resonance (the
unitarity region) a quantitative understanding of the phases comes mainly from
Monte-Carlo calculations~\cite{Carlson:2005kg}. (For other approaches
see~\cite{Radzihovsky:2007,Nishida:2006br,Randeria:2007}.) This is because 
close to  resonance  the scattering length is much larger than the inter-particle distance and 
 there is no small parameter in the Lagrangian to expand in. Therefore  fluctuations may change the mean field results substantially.  

In standard  BCS superfluids the chemical potentials of the two fermionic
species are equal.  An imbalance in the number of $\psi_1$ and $\psi_2$ is
implemented by taking the chemical potentials for the two species, $\mu_1$ and
$\mu_2$ respectively, to be different. (We will name our species in a way that
$\mu_1\geq\mu_2$.)  If the chemical potential difference,
$2\delta\mu=\mu_1-\mu_2$ is much smaller than the magnitude of the 
gap parameter $|\Delta|$, the splitting cannot disrupt BCS superfluidity because the
superfluid state with equal number densities is energetically favored in
comparison with a normal state with a fermionic imbalance. On the other hand, as
pointed out in \cite{Clogston:1962}, in the weak coupling regime, BCS
superfluidity cannot persist for large values of $\delta\mu$.  Indeed, there
exists an upper limit for $\delta\mu$ (the so-called Chandrasekhar-Clogston limit), beyond
which the homogeneous superfluid state is no longer energetically favored over
the normal phase.

 For imbalanced systems, a qualitatively complete picture of the phase diagram
has not been established yet. Proposed possibilities are phase-separation
\cite{Bedaque:2003hi}, breached pair superfluidity
\cite{Sarma,Liu:2002gi,Gubankova:2003uj,Forbes:2004cr}, deformed Fermi sea
pairing \cite{Muther:2002mc} and non-homogeneous or LOFF pairing \cite{LOFF}.
(See \cite{Casalbuoni:2003wh} and \cite{Yang:2006} for reviews.)

 The phase diagram of the system at $T=0$ as a function of the
scattering length and the chemical potential difference has been explored in the
mean field approximation in~\cite{Mannarelli:2006hr,Pao,Sheehy:2006qc,he}.  
The authors find that on the BCS side of the
resonance there are no stable homogeneous superfluid phases that have gapless
Fermi surfaces. On the BEC side of the resonance, there are stable gapless
superfluid phases, which can exhibit a net polarization. At resonance, mean
field theory suggests a first order phase transition from the superfluid to the
normal phase as $\delta\mu$ is increased, without any intervening gapless
superfluid phase.  Consequences of the phase diagram for experiments with
trapped atoms were explored in \cite{Sheehy:2006qc,trap}. At resonance if we
fill different number of $\psi_1$ and $\psi_2$ in the harmonic trap, because the
gapped phase can not feature a net polarization, the system phase separates with
an unpolarized superfluid  in the central region of the trap  and a polarized
normal fluid at the exterior.

For non-zero imbalance close to the resonance,  fluctuations  may  change the
mean field results qualitatively. This has to be contrasted with the zero
imbalance case,  where fluctuations lead only to a quantitative change of the
mean field results. Indeed for non-zero imbalance many features of the phase
diagram are not caught by the mean field approximation. The authors
of~\cite{Son:2005qx} go beyond mean field theory by using results from
Monte-Carlo simulations~\cite{Carlson:2007} and propose a  phase diagram which
features a splitting point near resonance at non-zero $\delta\mu$, where the
homogeneous superfluid, a LOFF like inhomogeneous phase, and the gapless
superfluid phase coexist. They also find stable gapless fermionic modes with one
and two Fermi surfaces, on the BCS side of the resonance. A detailed treatment
of fluctuations around the resonance using an expansion in $\epsilon = D-4$
space dimensions at $T = 0$ \cite{Nussinov:2006,Nishida:2006br} supports this
picture. A different approach consists in generalizing the Fermi gas to a model
with $2N$ hyperfine states, performing a systematic $1/N$ loop expansion around
the BEC-BCS solution~\cite{Nikolic:2007,Radzihovsky:2007}. The phase diagram at
unitarity has also been explored using a Superfluid Local Density Approximation
(SLDA) \cite{Bulgac:2006cv,Bulgac:2007wm}. With this method one finds that on
increasing $\delta\mu$ from zero at unitarity, there is an intervening window of
values for which the LOFF phase is favored over the homogeneous superfluid and
the normal phases.

In this paper, we study small fluctuations about the mean field value of the gap
parameter for a system with mismatched Fermi surfaces. We consider fluctuations
of $\Delta$  both in its phase and in its magnitude. Both of these involve 
a coherent change in the wavefunctions of fermions in many different momentum
eigenstates, and are therefore collective modes of the system. In particular, long wavelength fluctuations in
the Nambu-Goldstone field are associated with the hydrodynamic mode (or sound
mode) in the paired system and can be related to dynamic phenomena like
compressions in a trapped atomic gas~\cite{Dalfovo:1999zz}. By looking at the
stability of the energy with respect to these excitations, we can map out the
parameter values for which BCS-like pairing is favoured.  We also use the
expansion of the free energy in the magnitude of $\Delta$ to explore the typical
length scale of inhmogeneities in the condensate in non-uniform configurations
like vortices.

\section{Methods and materials}

In our quest to understand how fluctuations in the condensate about the mean
field value affect the phase diagram of cold atomic gases with unequal number of
$\psi_1$ and $\psi_2$ fermions, we study the effective Lagrangian density
describing these fluctuations.  We do this by integrating out the fermions from
the system and writing the effective action as a series in powers of the
fluctuations and their derivatives~\cite{Engelbrecht:1997,Randeria:2007}. We
explicitly calculate the terms up to second order in the fluctuations and their
derivatives. We expect that our mean field calculation of these coefficients are
under better control away from unitarity~\cite{Dalfovo:1999zz}.

For zero imbalance, the collective modes associated with fluctuations in
the phase and the magnitude of the condensate were analyzed over the full BCS-BEC
crossover in~\cite{Combescot:2006zz}. In the limit of long wavelengths (or small
momenta) the theory is dominated by the Nambu-Goldstone mode associated with the phase
fluctuations, travelling with the speed of sound given by $c_s^2 =
(n/m)(d\mu/dn)$. The study by~\cite{Combescot:2006zz} writes the effective
Lagrangian to all order in derivatives, but only to the second order in fields.
Very recently, in~\cite{Schakel:2009} the effective Lagrangian describing
interaction terms between the Nambu-Goldstone mode and the Higgs mode were
obtained. By integrating out the Higgs mode, the expression of the speed of
sound first obtained in~\cite{Marini:1998} was reproduced
in~\cite{Schakel:2009}.

In our study we restrict ourselves to only terms upto the second order 
in a derivative expansion. We reproduce the results of~\cite{Combescot:2006zz,Schakel:2009} 
and extend the analysis to non-zero imbalance. This is
a physically interesting case because experiments have been performed for
unequal number of $\psi_1$ and $\psi_2$, and a change in the behavior
of the collective modes can possibly give us information about novel phases that
may arise in these experiments. In particular, we find that the Higgs mode mass
shows an intersting behavior in the gapless BEC region as we discuss below. 
Because of this, we do not integrate out the Higgs mode as done
by~\cite{Schakel:2009}, and keep it in the effective Lagrangian.

Efforts to study the collective modes beyond the mean field approximation,
by methods that may be under better control near unitarity, can be found 
in~\cite{Son:2005qx,Radzihovsky:2007,Rupak:2007vp,Kryjevski:2007au}.

The coefficients of the terms in the effective action tell us about the
stability of the mean field solutions. The analysis of the stability of various
phases in imbalanced Fermi gases has been studied previously in several
different works. In~\cite{Sheehy:2006qc} the authors looked at the phase diagram
in detail, both in the narrow and the broad resonance limits. One
important conclusion from their study is that it is important to check that the
free energy is a local minimum rather than a local maximum, at the solution of
the gap equation. In Ref.~\cite{Pao:2006} it is  shown that this criterion is
equivalent to the requirement that the number susceptibility is positive. In
terms of the coefficients in the effective Lagrangian, it corresponds to the
requirement that the mass-squared of the Higgs field be positive, ensuring
stability with respect to homogeneous fluctuations. The authors
of~\cite{Sheehy:2006qc} also derived the Ginzburg-Landau theory in the BEC
regime for imbalanced Fermi gases, upto $\Delta^6$ in the fluctuations for the
Higgs field about the normal phase ($\Delta=0$). The motivation for considering
a Ginzburg-Landau expansion is that the gap is zero in the normal phase and
expected to be small close to the gapless superfluid-normal phase boundary. This
Ginzburg-Landau expansion can therefore be used to map the phase boundary
between the two phases~\cite{Sheehy:2006qc}. Our expression for the quadratic
coefficient in a Ginzburg-Landau expansion (shown in~\ref{ginsburg-landau}) can
not be directly compared to the expression in~\cite{Sheehy:2006qc} since this
specific expression was given only in the narrow resonance approximation, while
we work in the mean field approximation. However, by considering the stability
of the Higgs field we conclude that there are locally stable gapless phases in
the BEC regime, which go to the normal phase as we increase $\delta\mu$. This
conclusion matches the conclusion by~\cite{Sheehy:2006qc}. We also go further by
looking at the lowest non-trivial terms in the derivatives of the Higgs field.

 Several groups have analyzed stability with respect to space-time
dependent (inhomogeneous) fluctuations in the condensate. This can give
additional information to the study of local and global instability of
homogeneous condensates because a phase could be stable with respect to a
homogeneous change in the order parameter, but could be unstable with respect to
the formation of inhomogeneous condensates.

More specifically, the instability of gapless states towards the growth of phase
modulation of the condensate, the so called current instability, has been
studied in~\cite{Pao,Gubankova:2006gj,Chen}.  The instability towards a growth
of change in the magnitude (which will happen if the Higgs mass is imaginary),
the so called Higgs instability, was also studied by the authors of~\cite{Chen}.
They showed that the absence of this instability is equivalent to the
requirement that the number susceptibility matrix is positive
definite~\cite{Pao,Gubankova:2006gj}. It was also found that the  current
instability is much less stringent than the Higgs instability. (For the
manifestation of the current instability in the context of pairing in quark
matter, see~\cite{Huang:2004bg,Casalbuoni:2004tb,Fukushima:2006su}. The Higgs
instability in the quark matter context has been studied in~\cite{Gatto:2007ja,
Giannakis:2006vt,Giannakis:2006gg}).

By looking at the constraints on the positivity of the coefficients of the
effective action, we reproduce the above mentioned results for stability. In
addition, we consider the implication of the requirement that the energy cost of
creating a position dependent fluctuation in the magnitude of the condensate
(Higgs elasticity) be positive. This criterion has not been analyzed before in
the literature, but we find that it gives a weaker condition than current
stability.

The main new results in the present paper are related to a study of the Higgs mass and
Higgs elasticity as a function of the coupling and the chemical potentials. We
find that the Higgs mass is small in the gapless phase in the BEC regime.  We
also find that in the BEC regime, the Higgs elasticity is constant in the gapped
phase and increases in  the gapless phase. This has important consequences for
any non-homogeneous configuration created in a system tuned to sit in this
region. It implies that a cost of creating a gradient in the condensate value is
large, and hence the condensate should vary slowly in any such configuration. An
inhomogeneous configuration has been considered by~\cite{Silva}, who however
evaluaed the elasticity of the condensate field at unitarity  for vanishing
values of the gap, while we consider fluctuations about the mean field solution.
Our analysis is also an improvement over the analysis of~\cite{Lamacraft} where
the Higgs elasticity is not computed microscopically.

 This paper is organized as follows. In Section \ref{Sec-Model}, we present our
model and review some basic equations of the mean field analysis. In Section
\ref{Sec-Fluctuations}, we study the fluctuations of the difermion condensate,
and derive the general expression for the effective action for the fluctuation
fields, valid up to second order in the fluctuations. We consider both 
fluctuations in the magnitude, and in the phase of the difermion condensate.  The reader
not interested in the calculational details may skip over to Section
\ref{Sec-low}, where we present the low energy effective theory for these modes.
We show the expressions of the coefficients that appear in the effective
Lagrangian  for arbitrary values of the temperature. From the sign of these
coefficients we obtain stability criteria that we  analyze in detail  in the
case of vanishing temperature. 
From this analysis we reproduce the conclusion that there exist stable gapless phases in the
BEC (strong coupling) regime at non-zero asymmetry.
The central results of the paper are discussed in  Section
\ref{Sec-Conclusion}, where we evaluate the mass of the Higgs mode in the strong
coupling regime and find that for certain values of the parameters in the
gapless region, this mode is light. This implies that the outer core of a vortex
in this region will be wider than in the gapped superfluid phase. Furthermore,
we  find that the elasticity of the Higgs mode sharply increases in the
gapless region on increasing the asymmetry. This also means that the radius of vortices will be
large in this  region. This effect could be experimentally
detectable in cold atoms experiments where the mismatch between the two species
can be tuned.

 In \ref{appendix-relation} we show the equivalence between   the
coefficients in the effective action of the Nambu-Goldstone mode and the screening
masses that are obtained by gauging the $U(1)$ symmetry.  Formally, the
equivalence may seem apparent from gauge invariance, but the explicit
demonstration of the same is non trivial, and therefore we include the
derivation in \ref{appendix-relation}. In
\ref{appendix-coefficients} we report some details of the calculation of the
coefficients appearing in the effective Lagrangian.

\section{Calculation}
\subsection{Model and ansatz}
\label{Sec-Model}
 We consider a non-relativistic  system consisting of  two species of fermions $\psi_1$ and
$\psi_2$ of equal mass $m$ but different chemical potentials $\mu_1= \mu+\dm$
and $\mu_2= \mu-\dm$, with $\mu$ being the average of the two chemical
potentials and $2\dm$ the difference between them. Defining  the field
$\psi=(\psi_1\,\psi_2)^T$, the Lagrangian density describing free fermions can
be written as,
\begin{equation}
{\cal{L}}_f= \psi^\dagger\bigl(i\partial_t - E({{\bf p}}) + \mu
+\dm\sigma^3\bigr)\psi\;,
\end{equation}
where $E({{\bf p}})={{\bf p}}^2/(2m)$, with ${{\bf p}}$  the momentum operator
$\nabla/i$. The energy of a free fermion relative to the average chemical
potential is conventionally indicated  by ${\xip}=E({{\bf p}})-\mu$.  We assume
that the Feshbach interaction between fermions of different species  can be
modeled by a point like four Fermi interaction, and the corresponding term in
the Lagrangian can be written as
\begin{equation}
{\cal L}_I= \frac{\lambda}{2}\psi^\dagger_\alpha(x)\psi^\dagger_\beta(x)\psi_\beta(x)\psi_\alpha(x)\label{four fermi}\;,
\end{equation}
with $\alpha,\beta \in \{1,2\}$ and where $\lambda> 0$  for attractive interaction, the
case we are interested in.

 The effect of the attractive interaction between fermions is to produce a
difermion condensate
\begin{equation}
\langle\psi_\alpha(x)\psi_\beta(x)\rangle =
\frac{\Delta(x)}{\lambda}\varepsilon_{\alpha\beta}\label{condensate}\;,
\end{equation}
where $\varepsilon$ is the two dimensional antisymmetric tensor
$\varepsilon=i\sigma^2$.

In the mean-field approximation the Lagrangian density can be written  as,
\begin{equation}
{\cal{L}} = \Pd \left(
\begin{array}{cc}
i{\partial_t}-\xip+\dm\sigma^3 & -\Delta(x)\varepsilon    \\
\Delta^*(x)\varepsilon &   i{\partial_t}+\xip-\dm\sigma^3
\end{array} \right)\Psi -\frac{|\Delta(x)|^2}{\lambda}\label{Lagrangianmeanfield}\;,
\end{equation}
where $\Psi$ stands for the four component Nambu-Gorkov spinor,
\begin{equation}
\Psi = \frac{1}{\sqrt{2}}\left(\begin{array}{c}
 \psi_1\\
 \psi_2\\
 \psi_1^*\\
 \psi_2^*
\end{array}
\right)\label{define nambu gorkov}\;.
\end{equation}
The fluctuations of the condensate will be treated in the next
Section.  Here we only discuss the homogeneous phase, with
$\Delta(x)=\Delta=$ const. In this case the excitation spectrum is
described by the quasiparticle dispersion laws \be \epsilon_+ =
+\delta\mu + \sqrt{{\xip}^2+\Delta^2} \,,~~~~~~~~ ~~~~\epsilon_- = -\delta\mu    + \sqrt{{\xip}^2+\Delta^2} \label{dispersions} \, .\ee
The knowledge of  the dispersion laws of the system allows  one to evaluate the
grand-potential, which is given at $T=0$ by the expression,
\be \Omega_s -\Omega_n = \frac{\Delta^2}{\lambda}
-\frac{1}{2}\int \frac{d^3 p}{(2 \pi)^3} \, \Big[|\epsilon_+| +
|\epsilon_-| - 2 \xi({\bf p}) \Big] \label{omega}\, .\ee The integral in this
expression is ultraviolet divergent and can be regularized in the
usual way~\cite{Melo},
by writing $\lambda$ in terms of the scattering length $a$ according to \be \frac{m}{4\pi a} = \frac{1}\lambda + m \int
\frac{d^3p}{(2\pi)^3} \frac{1}{p^2} \, . \label{slength} \ee For later convenience we
introduce the dimensionless
 coupling constant \be g = \frac{1}{ k_F a}\, ,\ee
where  $k_F$ is the Fermi momentum of the system
which is defined in terms of the average number density $n$ of the two species
by the relation $n=k_F^3/(3\pi^2)$. The weak coupling regime, where
the BCS approximation holds, corresponds to $g \to -\infty$. This
approximation is generally very good for superconductivity in
metals. On the other hand, in cold atoms the strength of the
interaction can be varied in the vicinity of a Feshbach
resonance, where the scattering length strongly depends on the
applied magnetic field. Therefore both the weak and strong coupling
regimes can be reached in this case.

Knowing the free energy of the system, one can evaluate the gap parameter $\Delta$  
by solving the equation  \be
\frac{\de \Omega}{\de \Delta} = 0 \label{gap} \, . \\
\ee
Let us note explicitly that we do not write equations for $\mu_1$
and $\mu_2$. We do not work at fixed particle number densities $n_1$ and $n_2$ and therefore
we  do not impose the equations: \be \frac{\de
\Omega}{\de\mu_1} = - n_1 \qquad \frac{\de
\Omega}{\de\mu_2} = - n_2 \label{denneeq}\, ,\ee
which would be needed in the analysis if $n_1$ and $n_2$ were held fixed
\cite{Pao}. Instead, the values of $n$'s for given $\Delta$ and
$\mu$'s can be determined by the relations, Eq.~(\ref{denneeq}).

Note also that the effect of the condensate is to spontaneously break the global
$U(1)$ symmetry corresponding to the conservation of the  total  fermion number,
$n_1+n_2$. Therefore there will be a Nambu-Goldstone mode associated with the
spontaneous breaking of this  symmetry, and the system will consequently be a
superfluid. Clearly if one gauges this symmetry, the spontaneous breaking of the
local symmetry leads to the appearance of a mass term for the  gauge boson (Meissner mass), and the
system becomes a superconductor.  In the following analysis we will assume that
the  $U(1)$ symmetry  is global, i.e. fermions are not charged, and therefore we will study the dynamics of the
associated Nambu-Goldstone boson. In \ref{appendix-relation} we will
consider the relations between the parameters appearing in the Lagrangian
describing the Nambu-Goldstone bosons, and the screening masses of the gauge field.

\subsection{Fluctuations}
\label{Sec-Fluctuations}

In order  to include fluctuations of the condensate, we introduce the field
$\eta(x)$ that represents the deviation of the condensate from its mean field
value. In the presence of fluctuations, $\Delta(x)$ in Eq.
(\ref{Lagrangianmeanfield}) is given by
\be
\label{fluct}
\Delta(x) = \Delta+\eta(x) \,,
\ee
where it is assumed that the fluctuation is much smaller than $\Delta$.  In this
paper, we will consider only homogeneous condensates, meaning that $\Delta$ on
the right hand side of Eq.~(\ref{fluct}) is independent of $x$. In principle one
might consider the case where the underlying condensate is $x$ dependent, like
in the non-homogeneous LOFF phase. However, we will postpone the study of such a
case to future work.  In order to simplify the analysis, but without lack of
generality, we  choose the phase of the fermion fields so that the mean field
condensate, $\Delta$, is real. The field  $\eta$, on the other hand will have
both real and imaginary components.

For a given temperature $T$, the partition function is given by,
\begin{equation}
Z=\int{\cal{D}}\eta^*{\cal{D}}\eta{\cal{D}}\Pd{\cal{D}}\Psi
e^{-{\cal{S}}[\Pd,\Psi,\eta,\eta^*]}\;,
\end{equation}
where ${\cal{S}}$ is the Wick rotated action,
\begin{equation}
\begin{split}
{\cal{S}}[\Pd,\Psi,\eta,\eta^*] =&\intspace{x}\Bigl\{\frac{1}{\lambda}|\Delta+\eta(x)|^2
\\
&-\Pd \left(
\begin{array}{cc}
{-\partial_{x^4}}-\xip+\dm\sigma^3 & -(\Delta+\eta(x))\varepsilon    \\
(\Delta+\eta^*(x))\varepsilon  &   {-\partial_{x^4}}+\xip-\dm\sigma^3
\end{array} \right)\Psi\Bigr\}\;,
\end{split}
\end{equation}
and we use the imaginary time formalism where $x_4$ is the imaginary time $it$, and runs from $-1/(2T)$ to $1/(2T)$.

 To find the effective action for the $\eta$ field, we integrate out the
fermionic field, which can be done  because the action is
quadratic in $\Psi$. This gives,
\begin{equation}
Z=\int{\cal{D}}\eta^*{\cal{D}}\eta e^{-{\cal{S}}[\eta,\eta^*]}\;,
\end{equation}
with
\begin{equation}
\begin{split}
S[\eta,\eta^*] =& \intspace{x}\Bigl\{\frac{1}{\lambda}|\Delta+\eta(x)|^2\Bigr\}
\\
&- \Bigl\{\frac{1}{2}\tr\log \left(
\begin{array}{cc}
{-\partial_{x^4}}-\xip+\dm & -(\Delta+\eta(x))    \\
-(\Delta+\eta^*(x))  &   {-\partial_{x^4}}+\xip+\dm
\end{array} \right)
+(\dm\rightarrow -\dm)\Bigr\}\label{Seff1}\;,
\end{split}
\end{equation}
where  $\tr$ symbolizes the trace over  Nambu-Gorkov indices and over a complete set of functions over space-time.
The factor of $1/2$ before the $\tr$  takes care
of the fictitious doubling of degrees of freedom that arose when we
introduced the Nambu-Gorkov spinor.

 At a formal level, Eq.~(\ref{Seff1}) gives the desired effective action for the
fluctuations. However, it is not possible to compute  the $\tr$ analytically for
arbitrary functions $\eta(x)$ and hence we
expand the logarithm in increasing powers of $\eta$ (and $\eta^*$),
\begin{equation}
\tr \log(\hat{O}+\hat{V}) = \tr\log(\hat{O})+ \tr\Bigl(\sum_{n=1}^{\infty}
\frac{-1}{n}(-\hat{O}^{-1}\hat{V})^n\Bigr)\label{expansion}\;,
\end{equation}
where we have defined
\begin{eqnarray}
\hat{O} &= \left(
\begin{array}{cc}
A(p) & -\Delta    \\
-\Delta  &   {\tilde{A}(p)}
\end{array} \right)\;,\;
\hat{O}^{-1} &= \frac{1}{D(p)}\left(
\begin{array}{cc}
\tilde{A}(p) & \Delta    \\
\Delta  &   A(p)
\end{array} \right)
\label{define O} \\
\hat{V} &= \left(
\begin{array}{cc}
0 & -\eta(x)    \\
-\eta^*(x)  &   0
\end{array} \right) \label{define V}\;,
\end{eqnarray}
and where
\begin{equation}
A(p)=ip_4-{\xip}+\dm\,, \qquad 
\tilde{A}(p) = ip_4+{\xip}+\dm\label{define A}\;,
\end{equation}
with $p =(-\partial_{x_4},\nabla/i)$  the (Euclidean) four momentum
operator. The quantity appearing in the denominator of Eq.~(\ref{define O})
is given by \begin{equation}
D(p) \equiv A(p)\tilde{A}(p) -
\Delta^2=\bigl(ip_4+\dm+ \epsilon(\vp)\bigr)\bigl(ip_4+\dm-\epsilon(\vp) \bigr)\;,
\end{equation}
where we have also defined  $\epsilon(\vp)=\sqrt{{\xip}^2+\Delta^2}$.

  We thus obtain the effective action as a series expansion
\begin{equation}
\label{seriesS}
{\cal{S}}[\eta,\eta^*]={\cal{S}}^{(0)}+{\cal{S}}^{(1)}+{\cal{S}}^{(2)}+...\;,
\end{equation}
with ${\cal{S}}^{(i)}$ proportional to the $i$th power of $\eta$ (and $\eta^*$).
We shall now analyze the various terms in this expansion individually.

The  zeroth order contribution to the action ${\cal{S}}^{(0)}$ is
proportional to the free energy of the system in the absence of
fluctuations,

\begin{eqnarray}
{\cal{S}}^{(0)}  &=&(V/T)\Omega 
=(V/T)\Bigl\{ \frac{1}{\lambda}\Delta^2
-\Bigl[\frac{1}{2}\frac{T}{V}\sum_{p}\log(D(p))+(\dm\rightarrow -\dm)\Bigr]\Bigr\}\label{S0}
\;,\end{eqnarray}
where $V$ is the spatial volume of the system and $\Omega$ is the free energy at
finite temperature.  In Eq.~(\ref{S0}) and below, we will use a notation where
the sum over $p$ refers to a sum over (spatial) momentum eigenvalues, $\vp$, and
a sum over $p_4$ which runs over the fermionic Matsubara frequencies
$\omega_n=(2n + 1) \pi T$, for $n$ integer. Bosonic Matsubara frequencies,
$\omega_n=2n\pi T$,  will be  denoted with  $k_4$ .

Extremizing the free energy with respect to $\Delta$
we find two stationary points corresponding to the trivial solution $\Delta=0$, and
\begin{equation}
\frac{1}{\lambda} +\left[\frac{1}{2}\frac{T}{V}\sum_{p}
\frac{1}{D(p)} + (\dm\rightarrow-\dm)\right]=0\label{gap equation}\;.
\end{equation}
In the following we will assume that $\Delta$ is non-zero and use
Eq.~(\ref{gap equation}) to simplify various expressions.

We now turn to the term  of the action in Eq.~(\ref{seriesS}) that is linear in
$\eta$, i.e. ${\cal{S}}^{(1)}$. This term is given by,
\begin{eqnarray}
{\cal{S}}^{(1)} &=&
\frac{1}{\lambda}\intspace{x}\Bigl\{\Delta(\eta(x)+\eta^*(x))\Bigr\} -
\Bigl\{\frac{1}{2} \tr({\hat{O}}^{-1}{\hat{V}}) + \mdm \Bigr\}\nonumber\\
&=&  \Bigl\{\frac{\Delta}{\lambda}
 - \bigl[\frac{1}{2}\frac{T}{V}\sum_{p}
\frac{-\Delta}{D(p)} + \mdm \bigr]\Bigr\}
(\tilde{\eta}(0) + \tilde{\eta^*}(0))\;,
\end{eqnarray}
 where $\tilde{\eta}(k)$ and $\tilde{\eta^*}(k)$ are the Fourier transforms of
$\eta(x)$ and $\eta^*(x)$ and are given by
\begin{eqnarray}
\tilde{\eta}(k) &=& \intspace{x} \eta(x) e^{ik\cdot x}\nonumber\\
\tilde{\eta^*}(k) &=& \intspace{x} \eta^*(x) e^{ik\cdot x}\;.
\end{eqnarray}

 Employing the gap equation (Eq.~(\ref{gap equation})), one obtains that
${\cal{S}}^{(1)} = 0$. This is clearly  a consequence of the fact that we are
considering a  stationary point of the action.  This result also holds if
we  consider the solution $\Delta=0$.

 The lowest order non-trivial term in the expansion of the action  is 
the one quadratic in $\eta$
\begin{equation}
\begin{split}
{\cal{S}}^{(2)} =& \frac{1}{\lambda} \intspace{x}\Bigl\{\eta(x)\eta^*(x)\Bigr\} + \frac{1}{4} \tr
\Bigl\{({\hat{O}}^{-1}{\hat{V}})^2 + \mdm \Bigr\}\\
 =&\frac{1}{\lambda}
\frac{T}{V}\sum_k
\Bigl\{\tileta(-k)\tiletac(k)\Bigr\}\\
&+\frac{1}{4}\Bigl(\frac{T}{V}\Bigr)^2\sum_k\sum_p
\Bigl\{\frac{\Delta^2}{D(p)D(p+k)}
\bigl(\tiletac(-k)\tiletac(k)+ \tileta(-k)\tileta(k)\bigr)\\
&+\frac{2\tila(p)A(p+k)}{D(p)D(p+k)}
\tileta(-k)\tiletac(k)+ \mdm\Bigr\}\label{S2complex}\;,
\end{split}
\end{equation}
where the sum over $k$ means integration over the three-momentum $\bf k$ and sum
over bosonic Matsubara  frequencies.

Using the  the gap equation we can simplify the expression above as follows:
\begin{eqnarray}
{\cal{S}}^{(2)} &=&-\frac{T}{V}\sum_{k}\tileta(-k)\tiletac(k)\Bigl\{I_2(k)+2I_1(k)\Bigr\}
-\frac{T}{V}\sum_{k} \tileta(-k)\tiletac(k)
I_3(k)\nonumber\\
&&-\frac{T}{V}\sum_{k} (\tiletac(-k)\tiletac(k)+ \tileta(-k)\tileta(k))
I_1(k)\label{S2n}\;,
\end{eqnarray}
where we have defined,
\begin{eqnarray}
I_1(k)&=&\frac{-1}{4}\frac{T}{V}\sum_p\frac{\Delta^2}{D(p)D(p+k)}+
\mdm\nonumber\\
I_2(k)&=&\frac{1}{4}\frac{T}{V}\sum_p
\frac{(\tila(p+k)-\tila(p))(A(p+k)-A(p))}{D(p)D(p+k)}+ \mdm\nonumber\\
I_3(k)&=&\frac{-1}{4}\frac{T}{V}\sum_p\frac{\tila(p)A(p+k)-\tila(p+k)A(p)}{D(p)D(p+k)}+
\mdm\label{I_definition}\;.
\end{eqnarray}
Here $I_1(k)$, $I_2(k)$ and $I_3(k)$ are  even in $\vk$; $I_1(k)$ and $I_2(k)$
are even in the time component $k_4$ as well, while $I_3(k)$ is odd in $k_4$.
Therefore we have that $I_1(-k)=I_1(k)$, $I_2(-k)=I_2(k)$, and $I_3(-k)=-I_3(k)$
with $4$-d momentum $k$. Note that the ultraviolet  divergent contributions cancel exactly: $I_1$, $I_2$ and $I_3$ are all ultraviolet finite.

In order to clarify the expression that we have obtained, it is  convenient to 
separate $\eta$ into its real and imaginary parts,
\begin{equation}
\eta(x) =  \frac{1}{\sqrt{2}}(\lambda(x)+i\theta(x))\;.
\end{equation}
Thus, the  action ${\cal{S}}^{(2)}$ in Eq.~(\ref{S2n}) can be written in terms of the  $\lambda$
and $\theta$ fields, as
\begin{equation}
{\cal{S}}^{(2)}=
-\ha\frac{T}{V}\sum_k \left(\tilde{\lambda}(-k) \tilde{\theta}(-k)\right) \left( \begin{array}{cc}
I_2(k)+4I_1(k) & -iI_3(k)    \\
+iI_3(k) &    I_2(k)
\end{array} \right)
\left(\begin{array}{cc} \tilde{\lambda}(k) \\ \tilde{\theta}(k) \end{array} \right)\label{Action}\;.
\end{equation}
 The evaluation of the functions $I_1(k)$, $I_2(k)$ and $I_3(k)$ for
arbitrary values of $k$ is quite involved~\cite{Randeria:2007}. However, if we are interested in the
long wavelength fluctuations of the condensate, we can expand the integrals in
a power series in $k$ and obtain the low energy effective action of the system.

Note that for certain values of $\delta\mu$ and $\Delta$, the system may feature
gapless fermionic modes that  also contribute to the low energy dynamics of the
system~\cite{Mannarelli:2007bs}.

In the following Section we will study the low energy effective Lagrangian of
the system discarding the possible contribution of gapless fermions. This will
allow to elucidate the role  of the fields $\lambda$ and $\theta$.

\subsection{Low energy effective Lagrangian}
\label{Sec-low}

The physical meaning of    the real and complex components of the field $\eta$,
namely $\lambda$ and $\theta$, is easy to understand  in 
small fluctuation and long wavelength limit. We will show that  in this limit $\lambda$ corresponds
to the Higgs field and $\theta$ to the Nambu-Goldstone mode.

Moreover in the limit of small $k$ it is possible to expand $I_1(k)$, $I_2(k)$
and $I_3(k)$  in a power series  in $k$ and to evaluate analytically or
numerically each term of the expansion.

Upon making this expansion, we obtain to second order in $k$,
\begin{eqnarray}
I_2(k)&=& Ak_0^2-\frac{B}{3}\vk^2+{\cal{O}}(k^4)\nonumber\\
I_2(k)+4I_1(k) &=& - C + Dk_0^2-\frac{E}{3}\vk^2+{\cal{O}}(k^4)\nonumber\\
I_3(k) &=& -k_0 F + {\cal{O}}(k^3)\label{define ABCDEF}\;,
\end{eqnarray}
where the expressions of the coefficients  $A$, $B$, $C$, $D$, $E$ and $F$  are
reported in \ref{appendix-coefficients}. As a  check of our results we notice
that taking $\delta\mu=0$ in the expressions above, we reproduce the coefficient
of the effective Lagrangian obtained in~\cite{Schakel:2009}. In particular we
notice that for vanishing mismatch one has that $A=4 C$, which matches with the
result of~\cite{Schakel:2009}. However, for $\delta\mu \neq 0$ such a relation
does not hold.

To understand the physical meaning of the various coefficients in the effective
action, let us first consider the case  where the phase of the condensate, but
not its magnitude, fluctuates. That is,
\begin{equation}
\Delta \rightarrow \Delta e^{i\phi(x)}\label{phase fluctuations}\;.
\end{equation}
The field  $\phi(x)$ represents the  Nambu-Goldstone mode associated with the
spontaneous symmetry breaking of the total fermion number, $n_1+n_2$. Since
there is no term that explicitly breaks this symmetry, the mass of this
Nambu-Goldstone boson is exactly zero.

Then the   meaning of the coefficients $A$ and $B$ becomes clear if we notice
that to linear order in $\phi$, Eq.~(\ref{phase fluctuations}) corresponds to
$\lambda(x)=0$ and $\theta(x)=\sqrt{2}\,\Delta\,\phi(x)$. The low energy
Lagrangian density for $\phi$ is therefore,
\begin{equation}
{\cal{L}}_\phi = {\Delta^2} \bigl[A(\partial_t\phi(x))^2 -
\frac{B}{3}(\partial_i\phi(x))^2\bigr]\label{Goldstone}\;.
\end{equation}
Therefore  $A$ is the coefficient appearing in the kinetic energy density of the
Nambu-Goldstone mode, and $B$ is related to the spatial variation of the
Nambu-Goldstone mode.  A negative value of $B$ or of $A$ tells us that the mean
field solution of the system is unstable to the growth of  phase fluctuations of
the condensate. If both $B$ and $A$ are positive, the system is stable. In our
analysis we never find negative values of $A$, but we find regions of the
parameter space where $B$ is negative.

The speed of sound, or equivalently the speed of the Nambu-Goldstone mode, is
the same as the speed of $\phi$ field, $\sqrt{B/(3A)}$, in the weak coupling BCS
regime where integrating out the Higgs mode does not change the speed
significantly. Therefore we reproduce the well known weak coupling result. It is
easy to see that by integrating out the Higgs mode we reproduce the speed of the
Nambu-Goldstone mode calculated by~\cite{Dalfovo:1999zz} and verified
by~\cite{Schakel:2009}. Actually, from our expressions one can further extend
their result to non-zero $\delta\mu$.

Now consider the case where only the magnitude of the condensate fluctuates,
corresponding to the Higgs mode $\Delta\rightarrow\Delta+\lambda(x)/\sqrt{2}$.
The low energy Lagrangian density for these fluctuations is,
\begin{equation}
{\cal{L}}_\lambda = -\frac1 2 C\lambda(x)^2 + \frac1 2 D (\partial_t\lambda(x))^2 -
\frac{E}{6}(\partial_i\lambda(x))^2 \label{Higgs}\;,
\end{equation}
that for positive values of the coefficients $C$, $D$ and $E$ is equivalent to
the Lagrangian density of a massive bosonic field with mass squared (i.e. the square of
the gap in the excitation spectrum) equal to $C/D$. If
the various coefficients are not positive, then the system is unstable. We shall
now analyze the three terms appearing  in this Lagrangian.

The $C\lambda^2$ term  corresponds to the mass term and it can be interpreted as
the change in the free energy, reported in Eq.~(\ref{S0}), caused by
changing the magnitude of $\Delta$. Since the mean field value of $\Delta$ is
chosen so that the free energy is a local extremum, the sign of $C$ tells us
whether this extremum  is a local maximum, for $C<0$, or a local minimum, for
$C>0$. Therefore in the former case the system is unstable, in the latter it is
stable or meta-stable, depending on whether the local minimum
is also the absolute minimum of the system or not.
It can be shown analytically (and we have also checked  numerically)  that the curvature of the
potential around the stationary point is proportional to $C$.   If one of the
coefficient $D$ or  $E$ is negative, then  the mean field value is unstable with
respect to time- or space-dependent fluctuations of the magnitude of the
condensate. We find that  $D$ is always positive, whereas  $E$ is negative in a
certain region of parameter space.

Negative values of $B$ and $E$ are both related to the growth of
spatially non-uniform fluctuations of the condensate but may point
to different possibilities for the true ground state of the system.
A negative $B$ may suggest that the condensate prefers to develop a
non-zero phase modulation which carries a current, balanced by a
counter-propagating current carried by gapless fermions. A non-zero
$E$ points to the formation of a spatial modulation in the magnitude
of the condensate, which does not carry a current.

The coefficient $F$ does not appear in the discussions  of the
Nambu-Goldstone and Higgs Lagrangians above. $F$ mixes the $\lambda$ and $\theta$
components. Such mixing between the components of a complex field has been
discussed previously for Lagrangians featuring a global symmetry corresponding
to phase rotations of the
field~\cite{Kapusta:Thermal,Alford:2007qa}. In these cases the mixing term can be
interpreted as a chemical potential for the conserved charge. Here, the two
modes are not on an equal footing and the interpretation of this term may be
more involved. We leave further discussion of the mixing between the Nambu-Goldstone
and Higgs modes for future work.

Note that in  Eq.~(\ref{define ABCDEF}), the small momentum expansion of $I_1(k)$, $I_2(k)$ and $I_3(k)$ should be done with care. Namely, since in some cases these integrals
are divergent, one cannot interchange the order of taking small $k$ limit and $p$ integration.

%

\subsection{Analysis of stability at $T=0$.}
In the \ref{appendix-coefficients} we have reported the equations for the
coefficients  $A$, $B$, $C$, $D$ and $E$ for arbitrary values of the
temperature. However, in the present paper we content ourselves with the
analysis of the stability for the case of vanishing values of the temperature.

 Before considering the general case of arbitrary coupling, it is instructive to
consider the limiting case of weak interaction. At weak coupling, the BCS
hierarchy of scales, $\delta\mu, \Delta\ll \mu$, holds. Therefore one can carry
out the momentum integration analytically in a thin shell around the common
Fermi surface, $\mu$.

 Of particular interest is to study the phases which feature gapless fermionic
excitations. These phases  correspond to  $\delta\mu>\Delta$ and are
known to be unstable in weak coupling.  In this case, the
coefficients appearing in the Lagrangian of  the Nambu-Goldstone mode are
given by
\begin{eqnarray}
A&=&\frac{1}{8\pi^2}\frac{m(2m\mu)^{1/2}}{\Delta^2}\;(1-x)
\nonumber\\
B&=&-\frac{1}{8\pi^2}\frac{(2m\mu)^{3/2}}{m\Delta^2}\;\frac{1-x}{x} \; ,
\label{weak1}\end{eqnarray}
while  for the coefficients related to the  Higgs mode we obtain
\begin{eqnarray}
C&=&-\frac{1}{2\pi^2}m(2m\mu)^{1/2}\;\frac{1-x}{x}
\nonumber\\
D&=&\frac{1}{8\pi^2}\frac{m(2m\mu)^{1/2}}{3\delta\mu^2}\;\frac{1-x^3}{1-x^2}
\nonumber\\
E&=&-\frac{1}{8\pi^2}\frac{(2m\mu)^{3/2}}{3m\delta\mu^2}\;\frac{1-x^3}{x^3(1-x^2)} \; ,
\label{weak2}\end{eqnarray}
where we have introduced $x=\sqrt{\delta\mu^2-\Delta^2}/\delta\mu<1$. The mixing term is
\begin{eqnarray}
F=0 \,.
\end{eqnarray}
 The last equation shows that the Nambu-Goldstone and Higgs modes decouple in the weak
coupling.  Equations (\ref{weak1}) and  (\ref{weak2})  show that both the
Nambu-Goldstone and the Higgs fields develop instabilities in this regime, because the
coefficients  $B$,  $C$  and $E$ are negative.  The fact that  $B$
is negative indicates instability towards a phase with spontaneous generated
currents~\cite{Huang:2004bg,Casalbuoni:2004tb,Fukushima:2006su,Gatto:2007ja} and a negative
$E$ towards a modulation of the magnitude of the
condensate~\cite{Giannakis:2006vt,Giannakis:2006gg}.  Negative $C$ shows that
this gapless phase does not correspond to a local minimum of the energy.
However, in the weak coupling case it is known that well before the gapless
phase develops, there is a first order phase transition to the normal phase or
to a non-homogeneous superfluid phase. Indeed, for $\delta\mu > \Delta/\sqrt{2}$
the energy of the local minimum corresponding to the non trivial solution of the
gap equation is larger than the energy of the unpaired phase. This means that
the Higgs and the Nambu-Goldstone  modes that we are studying and that eventually become unstable at
$\delta\mu=\Delta$, correspond to fluctuations around the meta-stable solution
for $\delta\mu > \Delta/\sqrt{2}$.

\begin{figure}[h!]
\begin{center}
\includegraphics[width=3.in,angle=0]{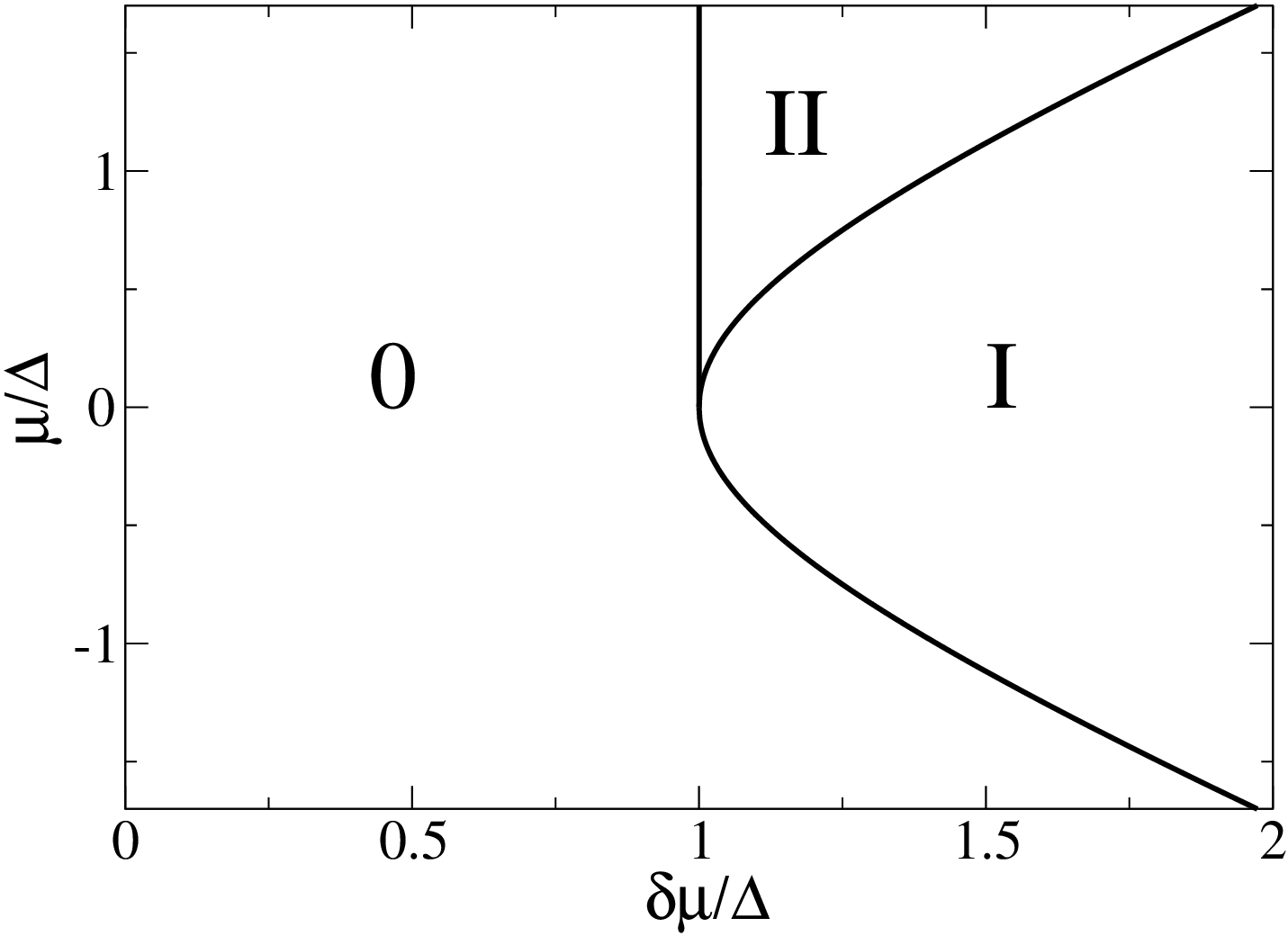}
\caption{Regions in the ($\bdmu$, $\bmu$)  plane which feature  zero, one or two
spherical surfaces in momentum spaces with gapless excitations. These regions
are marked with $0$, ${\rm{I}}$ and ${\rm{II}}$ respectively and are separated by solid lines.
 } \label{gapless regions}
\end{center}
\end{figure}

From this weak coupling analysis it is clear that in order to obtain
a stable gapless state one should study  the strong coupling regime
realized for larger values of the coupling constant.  To analyze the
stability of the various superconducting phases, we need to
calculate the values of $\Delta$ for given values of $\lambda$,
$\mu$ and $\delta\mu$ and then ascertain whether the coefficients
$A$, $B$, $C$, $D$ and $E$ are positive. In particular,  one of the
questions we are interested in from such a study is whether there
are regions of parameter space featuring stable phases having
gapless excitations on one or two spherical surfaces in momentum space.  One way to
study this question without  solving the gap equation is to
eliminate the variable $\Delta$ by writing $\mu$ and $\delta\mu$ in
units of $\Delta$~\cite{Gubankova:2006gj}. Therefore we define,
\begin{equation}
\bmu = \frac{\mu}{\Delta}\;\;,\;\;
\bdmu = \frac{\delta\mu}{\Delta}\;,
\end{equation}
and the coefficients $A$, $B$, $C$, $D$, $E$ and $F$  are
then functions of $\bmu$ and $\bdmu$, multiplied by appropriate powers of
$\Delta$ and $m$ to give the correct dimensions. We can then map out the region
in the ($\bdmu$,$\bmu$) space where the integrals are negative, indicating
instabilities.

 On the same ($\bdmu$,$\bmu$) plane  we can identify regions where the system
has gapless excitations. Of the two dispersion laws reported in
Eq.~(\ref{dispersions}) the one indicated with $\epsilon_-$  can become gapless
in a certain range of parameters. This dispersion law is given by
\begin{equation}
\begin{split}
\epsilon_-(\magp)&= -\dm + \sqrt{({\magp^2}/{(2m)}-\mu)^2+\Delta^2} =
\Delta\left(-\bdmu +\sqrt{(\bar{\magp}^2-\bmu)^2+1}\right)\\
&
{\mbox{with }}\;\;\;\bar{\magp}=\frac{\magp}{\sqrt{2m\Delta}}\;,
\end{split}
\end{equation} and it can have zeros as
a function of $\bar{\magp}$ (or $\magp$).

In Fig.~\ref{gapless regions} we have divided the ($\bdmu$, $\bmu$) plane in three
regions corresponding to the different number of gapless surfaces in momentum space and marked such
regions with  $0$, ${\rm{I}}$ and ${\rm{II}}$. The region marked with $0$
corresponds to $\bdmu<1$ or  $\bdmu>1$ with $\bmu<-\sqrt{\bdmu^2-1}$, where the
dispersion law $\epsilon_-$ has   zero gapless modes.  Region ${\rm{I}}$ corresponds to
$\bdmu>1$ and $\bmu\in [-\sqrt{\bdmu^2-1},+\sqrt{\bdmu^2-1}]$, where
$\epsilon_-$ is zero on one spherical surface in momentum space. Finally, the
region ${\rm{II}}$ corresponds to $\bdmu>1$ and $\bmu>+\sqrt{\bdmu^2-1}$ where
$\epsilon_-$ is zero for two distinct values of $\magp$, corresponding to two
spherical surfaces in momentum space.

In Ref.~\cite{Gubankova:2006gj} an analysis of the stability of  the various
 regions reported in this diagram has been done. In that paper the following
 requirements  have been considered:
\begin{description}
\item[\hspace{.5cm}i.]{The Meissner mass of two fictitious gauge bosons that couple to the
fermions $\psi_1$ and $\psi_2$ should be real and positive.}
\item[\hspace{.5cm}ii.]{The $2\times2$ number susceptibility matrix associated with the two chemical potentials $\mu_1$ and $\mu_2$ should be positive definite.}
\item[\hspace{.5cm}iii.]{The free energy of the superconducting state should be lower than the free
energy of the unpaired state, meaning that the pressure in the superconducting phase has to be larger than the pressure in the normal phase.}
\end{description}
It turns out that the positivity of the Meissner mass leaves some
region in the parameter space where the gapless state with two Fermi
surfaces is stable. However, requiring the positivity of
susceptibilities eliminates all the gapless states with two Fermi
surfaces.  Considering all the stability  criteria above leaves only
a narrow strip at $\mu<0$, where the gapless state with one Fermi
surface is  stable. 

We conduct a similar study by requiring that the coefficients $A$, $B$, $C$, $D$
and $E$ are positive. It turns out that $A$ and $D$ are
positive in all parameter space, while the other coefficients are negative in
some regions. Since $A$ and $D$ turn out to be  positive in the whole ($\bdmu$,
$\bmu$) plane, the requirements of stability can be expressed throught the following criteria:

\begin{enumerate}
\item{ The coefficients $B$ and $E$ must be positive. This corresponds to have a
real speed of sound for the Nambu-Goldstone mode and for the Higgs mode.}
\item{ The
coefficient  $C$ must be positive. This corresponds to requiring that the
superfluid state  is a local minimum of the free energy. }
\item{The free energy of
the superfluid state should be lower than the free energy of the unpaired state,
i.e. $\Omega_s-\Omega_n<0$, where $\Omega_s$ and $\Omega_n$ refer to the free
energies of the superfluid and the normal phases, respectively. }
\end{enumerate}

Notice that according to~\cite{Pao:2006} the stability criterion 2 is equivalent to criterion {\bf ii.} above.
The stability criteria 2 and 3 have been used to map out the phase diagram of
imbalanced Fermi gases at both zero and non-zero temperatures in
Refs.~\cite{Sheehy:2006qc,Pao,Pao:2006,Gubankova:2006gj,Mannarelli:2006hr}.
In these papers it is shown that the most stringent condition, for any values of $\mu$ and $\delta\mu$, is
that the free energy of the superfluid state should be lower than the free
energy of the unpaired state, corresponding to criterion 3 above.  
Here we want to remark that  the requirement that the coefficients
$B$, $C$ and $E$ are positive, does still give some information about the system.
Consider as an example gapless states that satisfy criteria 1 and 2, but fail 3. In this case the
 system is in a  metastable gapless states that may be realized and studied in experiments.

\begin{figure}[h!]
\begin{center}
\includegraphics[width=2.4in,angle=0]{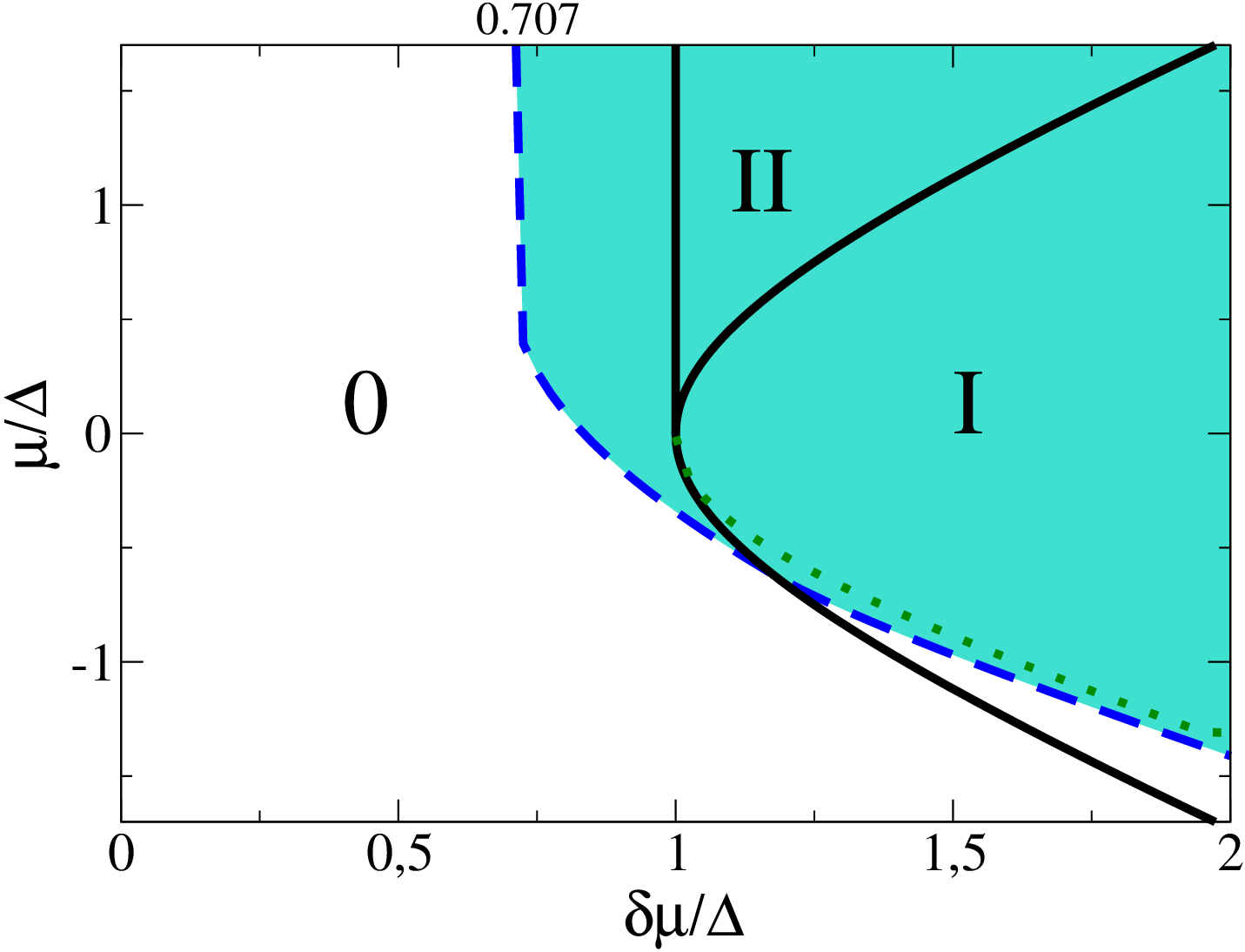}
\includegraphics[width=2.4in,angle=0]{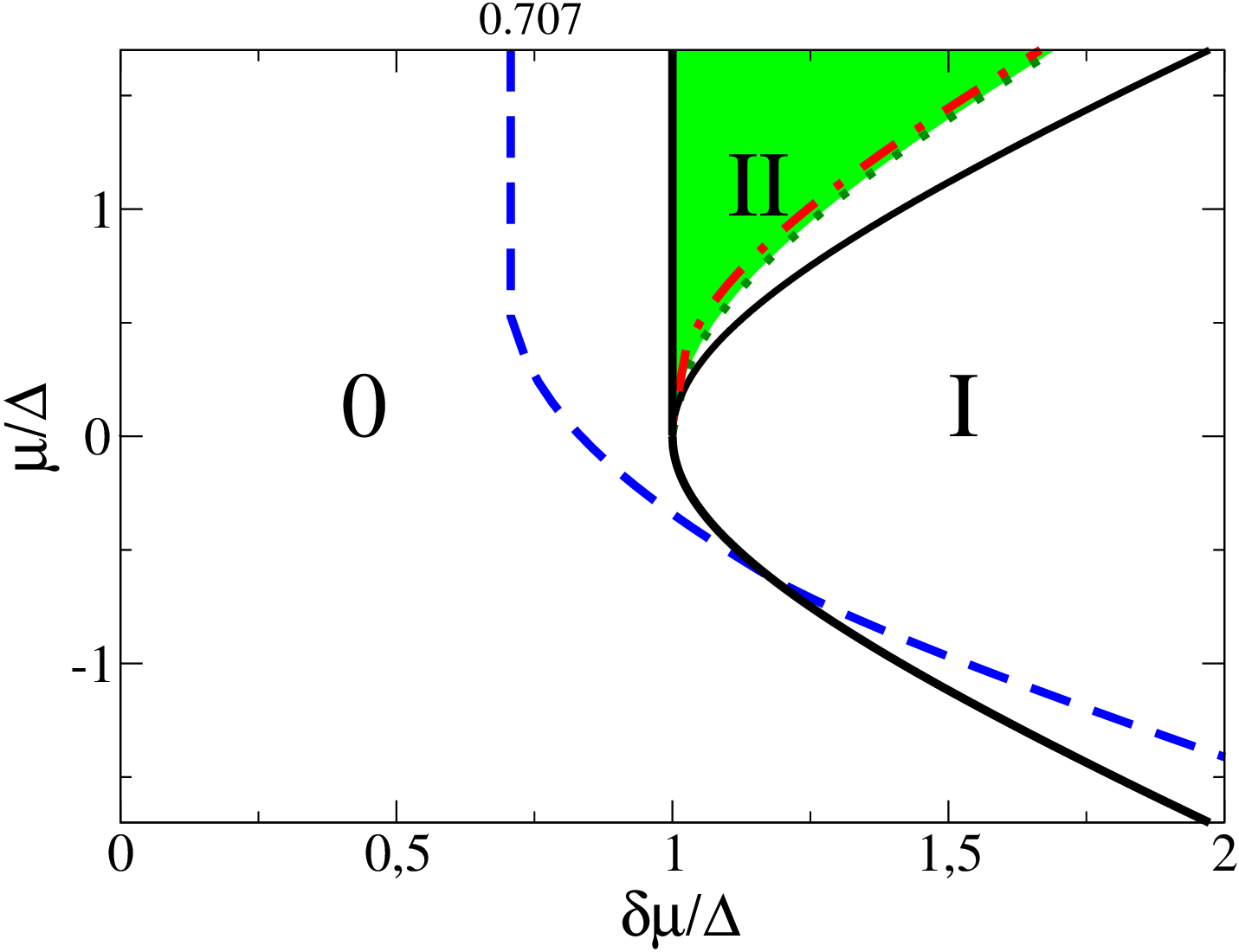}
\caption{(color online) Left panel: Stable and unstable regions  in the
($\bdmu$, $\bmu$) plane according to criteria  2 and 3. The requirement
$\Omega_s-\Omega_n>0$, (criterion 3) excludes the shaded region directly above
the dashed blue curve. We will refer this curve as ``curve 3'' below. The
requirement $C>0$ (criterion 2), excludes the region directly above the  green
dotted curve. We will refer to this curve as ``curve 2'' below. Criterion
3 is more restrictive than criterion 2; there  is a sliver of parameter space
where the superfluid phase corresponds to  a local minimum but not a global
minimum of the free energy. Right panel: Regions in the ($\bdmu$, $\bmu$) plane
which are stable or unstable according to the criterion 1 and 3. The
requirement $B>0$ excludes the shaded region  directly above the dotted green
line. The requirement $E>0$ excludes the region  directly above the  dot-dashed
red line. We see that the requirement $B>0$ is more restrictive than the
requirement  $E>0$. In any case, these  two requirements leave regions of
parameter space showing two gapless surfaces, however this region is excluded
once the criterion 3, corresponding to the dashed blue line, is considered. On
the top of both figures the Chandrasekhar-Clogston limit $\delta\mu/\Delta =
1/\sqrt{2}\simeq 0.707$ is indicated, which corresponds to the critical value of
the chemical potential splitting for the favorability of the superfluid phase in
weak coupling. $0$, ${\rm{I}}$ and ${\rm{II}}$ refer to the regions with zero, 
one and two gapless surfaces respectively, as in Fig.~\ref{gapless regions}} \label{C>0 region} 
\end{center}
\end{figure}

In Fig.~\ref{C>0 region} we report the results of our analysis concerning  the
stability criteria 1, 2 and 3 above.  On the left panel we report the results
regarding the stability criteria 2 and 3. Criterion 3, corresponding to
the requirement that $\Omega_s-\Omega_n>0$, excludes the shaded region directly
above the blue dashed line. Criterion 2, corresponding to the requirement $C>0$,
excludes all the region directly above the dotted green line.  A comparison
with~\cite{Gubankova:2006gj} shows that the requirement that $C>0$ is equivalent
to the condition that the number susceptibility of the system should be
positive.  Therefore there is a sliver of parameter space where the superfluid
phase is meta-stable and not absolutely stable. Our findings are in agreement
with the results of Ref.~\cite{Sheehy:2006qc}, where it is found that deep in
the BEC region a meta-stable gapless state exists.

While criterion 1, corresponding to the requirement that $B$ and $E$ are positive, is not
as restrictive as criterion 2,  it is still interesting because it tells us
about the tendency of the system to turn into a non homogeneous phase.  On the right panel of 
Fig.~\ref{C>0 region} we report the results of the stability analysis
concerning  the coefficients $B$ and $E$. The requirement $B>0$ excludes the shaded
region of parameter space directly above the dotted green line and is equivalent
to the requirement that the Meissner mass be real~\cite{Gubankova:2006gj}.
Requiring $E>0$ excludes the region directly above the dot-dashed red line.
Therefore, the requirement $B>0$ is more restrictive than the requirement  $E>0$.
Hence we find that the additional consideration of the position dependent
fluctuations in the Higgs field does not yield a more stringent criterion for
stability than the requirement that there be no current instability.

Notice that these criteria do not forbid the existence of states with  two
gapless surfaces. For reference, the dshed blue 
curve corresponding to the criterion 3 is also reported in the right panel of Fig.~\ref{C>0 region}.
We now look at the implications of the variation of the 
expansion coefficients as a function of $\delta\mu$, $\mu$ and $\Delta$, 
for the variation of the length scale of the modulation of the condensate 
in vortices.

\section{Results and discussion}
\subsection{Parameters of the Higgs Lagrangian and vortex radius}
\label{Sec-Conclusion}
 The requirement that small fluctuations in the magnitude and the phase of the
order parameter increase the free energy rather than decrease it, provides a
strong constraint on the values that $\Delta$, $\mu$ and $\dmu$ can take
in asymmetric cold atomic systems. The strongest constraint from these ``local
criteria'' comes from the requirement that the value of $\Delta$ be a local
minimum of the free energy rather than a local maximum (criterion 2). This condition excludes
the possibility that there can be two spherical surfaces in momentum space
featuring gapless quasiparticle excitations. A stronger constraint is provided
by a global condition that the homogeneous superfluid phase has a lower free
energy than the normal phase (criterion 3).

\begin{figure}[h!]
\begin{center}
\includegraphics[width=3.5in,angle=0]{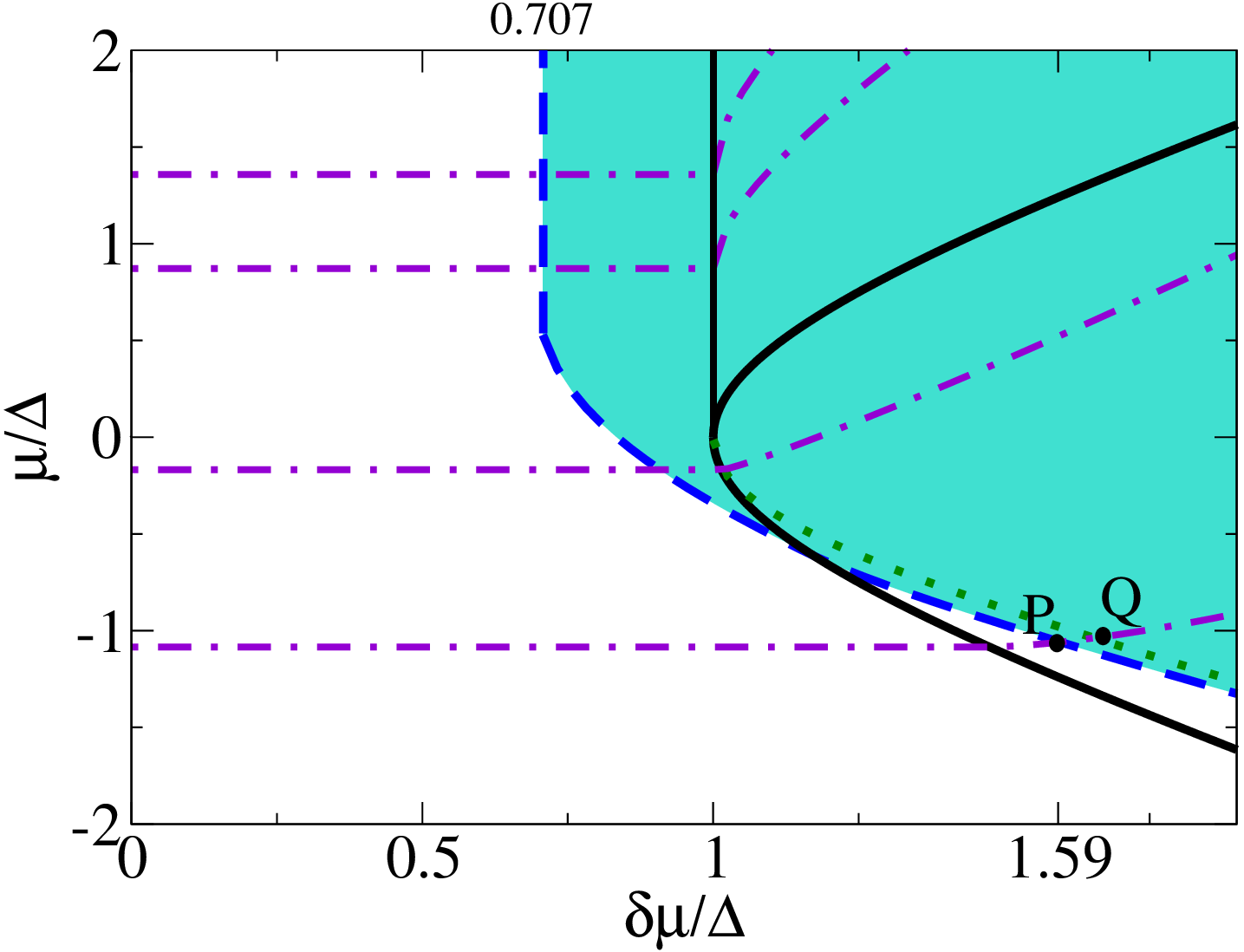}
\caption{(color online) Curves of constant $\kappa$ in the $\delta\bar{\mu}$, $\bmu$ plane.
Shown are four curves (dot-dashed lines, purple online) corresponding to
$\kappa=-0.5$, $\kappa=0$, $\kappa=1$ and $\kappa=1.71$. Negative scattering
lengths, and therefore negative $\kappa$, correspond to the BCS regime while
positive $\kappa$ to the BEC regime. We look at the Higgs mass as a function of
$\delta\bmu$ along the curve corresponding to $\kappa =1.71$ in Fig.~\ref{higgs
mass}. Point $P$, corresponding to $\bdmu \simeq 1.59$, has the largest value of $\delta\bmu$ along the curve, for
which $\Omega_s-\Omega_n<0$. Between $P$ and $Q$, corresponding to $1.59 <\bdmu < 1.66$,  the superfluid phase is metastable. To the right of the point $Q$, corresponding to $\bdmu > 1.66$, the superfluid phase
is locally unstable, meaning $C<0$.}
\label{contours}
\end{center}
\end{figure}

 From Fig.~\ref{C>0 region}, one can notice that for $\dmu>\Delta$, the curves
associated with criterion 2 and criterion 3 run very close in the
gapless region. Indeed these two curves appear to converge asymptotically,  for
$\delta\mu \gg \Delta $. We recall that the Higgs mass is zero along curve 2
(dotted line (green online) in Fig.~\ref{C>0 region}). This suggests that the
mass of the fluctuations in the magnitude of the condensate is very small along
curve 3 (dashed line (blue online) in Fig.~\ref{C>0 region}) in the region $I$,
and gets smaller as the two curves come closer. Since the presence of a light
Higgs mode may be experimentally detectable, we have  explicitly studied the
mass of the Higgs field in the region where $\mu<0$, as a function of $\dmu$.
This region in parameter space is accessible with positive values of the
scattering length $a$, and lies on the BEC side of the resonance.

To be concrete, we first solve the gap equation for various scattering lengths
and see where we land in the parameter space. The result is shown in
Fig.~\ref{contours}. The four dot-dashed lines (purple online) show how
$\bar{\mu}$ varies as a function of $\delta\bar{\mu}$ for four different values of
the dimensionless variable, $\kappa=\pi/({2\sqrt{2m\Delta}a})$~\cite{Gubankova:2006gj}.
Values of  $\kappa \ll -1$ correspond to being deep in the BCS regime,  while
$\kappa \gg  1$ corresponds to being deep in the BEC regime.  Since we are
interested in the  BEC  regime we consider, for definiteness, the curve
corresponding to $\kappa=1.71$. It intersects the curve corresponding to
criterion   3 in $P$, at   $\bdmu \simeq 1.59$,  and the curve corresponding to
criterion  2 in $Q$, at $\bdmu \simeq 1.66$. 

\begin{figure}[h]
\begin{center}
\includegraphics[width=3.in,angle=0]{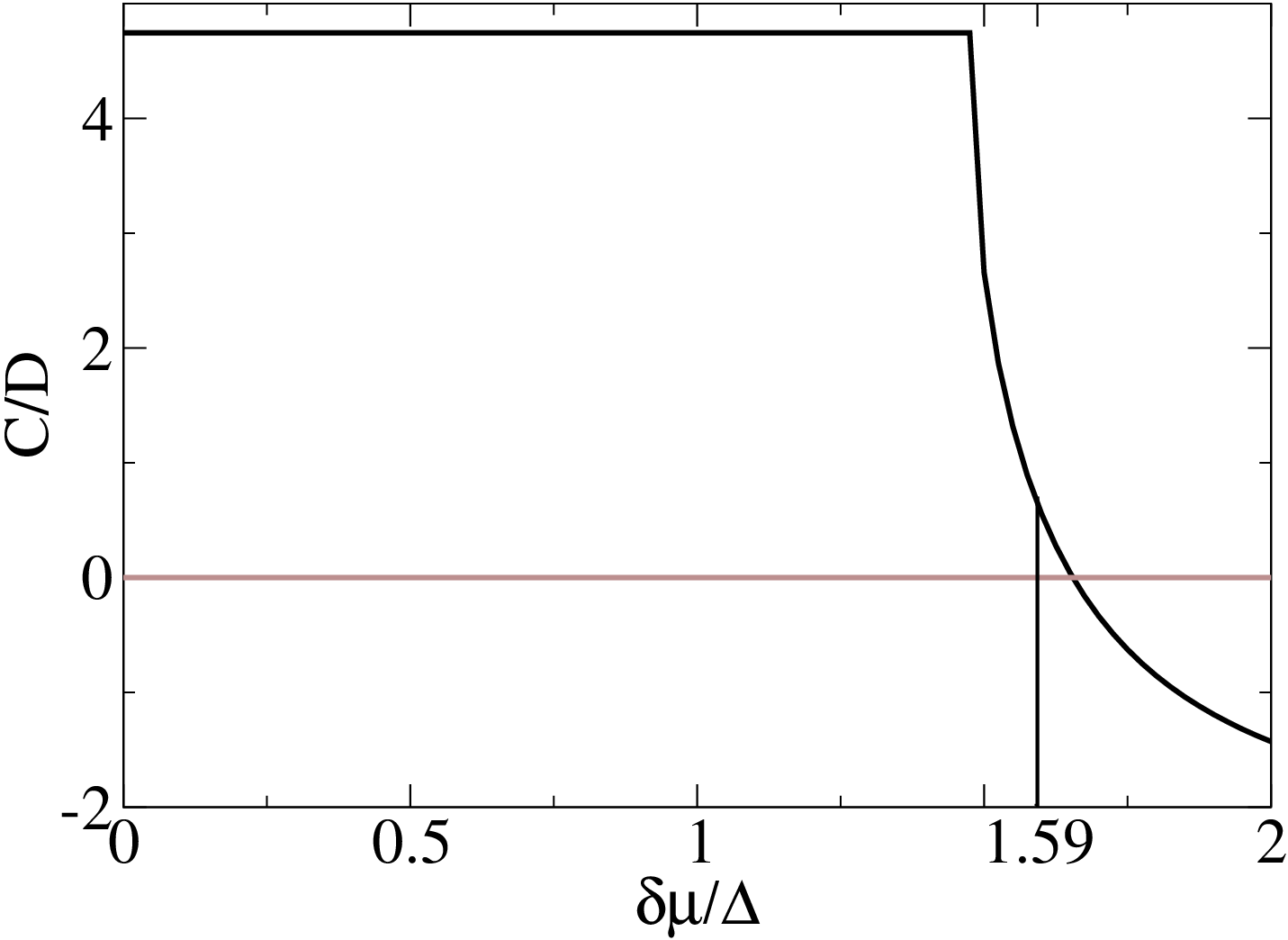}
\caption{(color online) Mass squared of the Higgs mode,   $m_H^2=C/D$, in units of $\Delta^2$,
as a function of $\bdmu$ along the curve $\kappa=1.71$ (see
Fig.~\ref{contours}). In the gapped region $m_H^2$ is a constant, and decreases
when we enter the gapless regime. At $\bdmu\simeq 1.59$ the Higgs mass has the
smallest value  in the regime where the homogeneous superfluid is favored over
the normal phase. For $\bdmu > 1.66$,  corresponding  to points on the right of
$Q$ in Fig.~\ref{contours}, the Higgs mass becomes imaginary and the superfluid
phase  is  locally unstable.} \label{higgs mass}
\end{center}
\end{figure}

Now consider the value of $C/D$, which is the mass squared of the Higgs
fluctuation of the condensate, as we increase $\delta\bar{\mu}$
along the curve labeled $\kappa=1.71$ in Fig.~\ref{contours}. (See
Fig.~\ref{higgs mass}.)
As can be seen in Fig.~\ref{higgs mass}, for
$\bdmu=0$, the superfluid phase is favored
over the normal phase and is also locally stable, meaning $C>0$. As
we increase $\bdmu$, as long as we are in the gapped phase, the free
energy of the superfluid phase is independent of $\bdmu$, (although
$\Omega_n$ decreases as we increase $\dmu$) and hence the mass
squared of the Higgs is  positive and independent of $\bdmu$ in this
region. As we cross into the region featuring one gapless surface,
$C$ decreases as we move closer to the curve 2. When $\bdmu \simeq 1.59$,
corresponding to point $P$ in Fig.~\ref{contours}, we have reached  the largest
value of $\bdmu$ for which the superfluid phase wins over the normal
phase. This gives the smallest value of the Higgs mass in the region
where it describes oscillations about the global minimum. We note
that $m_H^2$ drops by a factor of about $7.5$ at $\bdmu \simeq1.59$ from
its value at $\bdmu=0$.

 Moving along into the metastable region between point $P$ and $Q$, the Higgs
mass square decreases and finally becomes negative when we cross curve 2 at point $Q$ in
Fig.~\ref{contours},  corresponding to $\bdmu \simeq 1.66$.

Note that this calculation is done in a region where mean field methods are
expected to be reliable. To illustrate this, we calculate the value of the
inverse of the dimensionless expansion parameter, $g=1/(k_Fa)$.  Large and
negative values of $g$ correspond to being  deep in the BCS regime  while large
and positive values of $g$ correspond to the region deep in the BEC regime.   At
the point $P$ one has $g = 1.31$, and for larger values of $\kappa$, $g$
will be even larger, meaning a more
reliable predictions for the  mean field method.

 The low energy field theory describing a system tuned to be near  point $P$,
will have a very interesting particle content. It will consist of gapless
fermions living on one surface in momentum space, massless fluctuations in the
phase of the condensate and massive but very light fluctuations in the
magnitude of the condensate. It would be interesting to find some observable
that might be experimentally measured in order to probe such a spectrum.

 The fact that the mass is particularly small also implies that quantum
corrections may significantly alter its value. A renormalization group analysis with these
three degrees of freedom can clarify how beyond mean field corrections may shift its
value. We leave this for future work.

 But even before such a detailed study, we propose a striking consequence of our
results. The correlation length $r_0$, or the typical length scale at which the
magnitude of $\Delta$ varies in field configurations that arise when the system
is excited, is inversely proportional to the mass of the Higgs mode of the system.
For example, $r_0$ governs the size of the outer core of a vortex configuration
in a superfluid phase. This can be seen more concretely by writing the classical
field equations for a condensate of form
$\Delta({\bf{r}})=(\Delta+\rho(r))\exp(i\varphi(\phi))$, where $\Delta$ is the ground
state value of the condensate and $(r,\phi)$ are the
cylindrical polar coordinates with the vortex at $r=0$.
(See \cite{Dalfovo:1999zz} for reviews and references
therein.)  The boundary conditions for
the field $\rho(r)$ are that $\rho$ should tend to $-\Delta$ at the center of
the core (where the small fluctuation approximation begins to break down) and
should tend to $0$ as $r$ tends to infinity. For a vortex configuration,
$\varphi$
winds around by a multiple of $2\pi$ as we traverse a loop around the
vortex. Sufficiently far away from the inner core of the vortex, the spatial derivative of
$\varphi(\phi)$ does not contribute significantly to the equation of motion, and the classical field equation 
for static $\rho({\bf{r}})$ is,
\begin{equation}
\rho({\bf r}) - r_0^2 \nabla^2 \rho({\bf{r}}) =  {\rm{const.}}\label{eqn of
motion}\; ,
\end{equation}
where 
\begin{equation}
r_0 = \sqrt{E/(3C)}
\label{r0} 
\end{equation}
From Eq.~(\ref{eqn of motion}), it is clear that $\rho$ will decrease from a value close to $0$ to a
value close to $-\Delta$, as we go closer to the inner core of the vortex, over a
length scale $r_0$.  The fact that $C$ is numerically small close to the point 
$P$ in parameter space~(Fig.~\ref{contours}), will manifest itself in an increased size for the
outer core of the vortex.

This is admittedly a simplified discussion. For example, to construct an actual
vortex solution, it will be important to include the $\eta^4$ term in the
effective action. But the coefficient of this term is dimensionless, and would
not introduce any additional length scale in the problem, and hence we expect
our basic argument to remain valid in such a detailed study~\cite{note1}.

\begin{figure}[h]
\begin{center}
\includegraphics[width=3.in,angle=0]{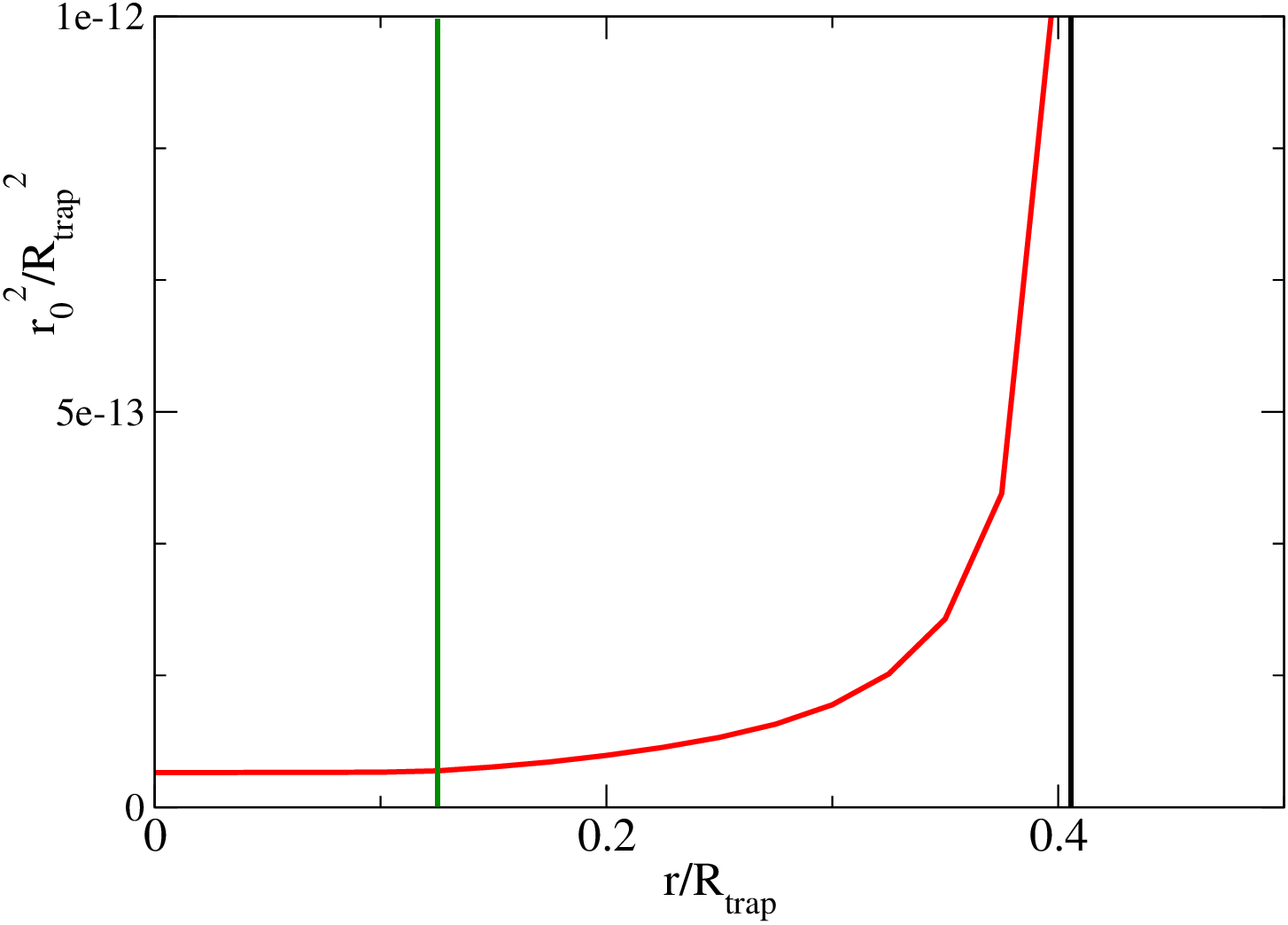}
\caption{(color online) Outer vortex radius as a function of the radial position
$r$ in a spherically symmetric trap at vanishing temperature. Distances are
scaled by $R_{\rm{trap}}$, which is the distance at which the effective
chemical potential becomes equal to $-1/(2ma^2)$, where $a$ is the scattering
length. The trap parameters are given in the text. On the right of the vertical
green line the excitations are gapless. On the right of the vertical black line
the system is in the normal phase. The value of the vortex outer radius changes
only very slowly in the gapped region and increases monotonically in the gapless
region. The divergent large value of the outer core vortex radius corresponds to
a transition to the normal state where vortex does not exist. Notice that the
vortex radius exhibits change in the derivative at the gapless point which may
serve as a signature for the gapless phase.}
\label{r0sq}
\end{center}
\end{figure}

To see the effect quantitatively, we plot in Fig.~(\ref{r0sq}), the outer vortex
radius square, $r_0^2$, as a function of a position away from the center of a
harmonic trap. We use a standard harmonic trap which models a potential in
optical lattices. The trap parameters used are $\omega=1.25\times10^{-13}$eV, which gives
for $m=5.61\times 10^{9}$eV for $Li$, a potential $m\omega^2
r^2/2=\omega(r/r_0)^2/2$ with $r_0=37.8$eV$^{-1}$ .
At the center of the trap $\mu=-8\times 10^{-7}$eV. The splitting $\delta\mu=1.15\times
10^{-6}$eV is
constant throughout the trap. $R_{\rm{trap}}$, the distance from the center at
which at which the effective chemical potential becomes equal to $-1/(2ma^2)$ is
then $4.6\times 10^{4}$eV$^{-1}$. We scale the distance by this radius. We want to be in
the BEC side and choose $a=1\times 10^{-2}$eV$^{-1}$. With these parameters, the
gap at the center of the trap is $8.29\times 10^{-7}$eV. The trap parameters are chosen as an
illustration of what effects can be seen by choosing a trap which has a
substantial volume in a gapless phase. 

As we go out from the centre, the effective chemical potential $\mu-V(r)$, and
therefore $\Delta$ decreases and at $r/R_{\rm{trap}}\sim 0.12$ (corresponding to the vertical green line in
Fig.~\ref{r0sq}) we move into the gapless regime. In the gapless region the
radius of the vortex increases monotonically until it formally diverges as we
enter into the normal state with no superfluid vortices.

The increase of the radius of the vortex core with increasing mismatch in the
gapless region can be qualitatively explained comparing the kinetic energy of a
superfluid element close to the superfluid vortex with the ``condensation
energy" associated with the superfluid phase; the condensation energy being the
difference between the free energy in the homogeneous phase and in the normal
phase. The definition of the vortex radius is by itself ambiguous, because there
is no abrupt transition from the superfluid phase  to the normal phase and
various definitions have been proposed, see {\it e.g.}~\cite{barenghi:1983}.
However the length scale at which the condensation energy is equal to the
kinetic energy should give a qualitatively correct result. In particular we
expect that the vortex radius estimated with this methods should increase
steeply in the gapless phase. 
The kinetic energy of a fluid element close to a vortex is given by  
\be \label{kin}
E_k = n \frac{1}2 m v^2 =  \frac{n}{8 m r^2} \,,
\ee
where $n$ is the local superfluid density,  $m$ is the mass of the atom and  the velocity of superfluid matter near a vortex is given by 
\be
{\bf v}(r) = \frac{1}{2 m r} {\bf e}_{\theta} \,,
\ee
with  $r$  the radial distance from the center of the vortex and   ${\bf e}_{\theta} $  the tangent unit vector.  As we approach the vortex core the velocity increases and consequently the kinetic energy increase. In principle the velocity and the kinetic energy diverges for $r \to 0$, signaling that a certain point, {\it i.e.} at a certain value of $r$,  a phase transition to the normal phase has to take place. 

The condensation energy is given by
\be \label{cond}
E_{\rm cond} = n \epsilon_{\rm cond}\,,
\ee
where $\epsilon_{\rm cond}$ is given by the difference between the free-energy densities of the superfluid phase and of the normal phase. Equating Eq.(\ref{kin}) to Eq.(\ref{cond}) we find that the vortex radius is given by
\be\label{radius}
\tilde r_0 = \sqrt{\frac{1}{ 8 m \epsilon_{\rm cond}}} \,.
\ee
In the gapped region the energy difference between the superfluid phase and the
normal phase is not strongly dependent on $\delta\mu$, thus $\tilde r_0$ is
approximately constant. In the gapless phase the  condensation energy
continuously decreases on increasing asymmetry and tends to zero at the boundary
between the gapless and the normal phase.  Thus, $\tilde r_0$ continuously
increases in the gapless phase and   at the boundary between the gapless phase
and the normal phase $\tilde r_0$ diverges. 

Notice that this definition of the radius of the vortex core has to be taken
with care, because in the normal phase the condensation energy vanishes and
Eq.~(\ref{radius}) seems to suggest that $\tilde r_0$ diverges. However, in the
normal phase there is no superfluid motion, consequently there are no superfluid
vortices.

The properties of vortices in the gapless region can not be tested in the BCS
regime however because there is no stable gapless region in the BCS side. Therefore we
look at the BEC regime. Qualitative explanation of why the vortex size increases
with the mismatch, Eq.~(\ref{radius}), can not be applied to the BEC region with
negative chemical potential, $\mu<0$. Below we consider another approach.   

\begin{figure}[h]
\begin{center}
\includegraphics[width=2.3in,angle=0]{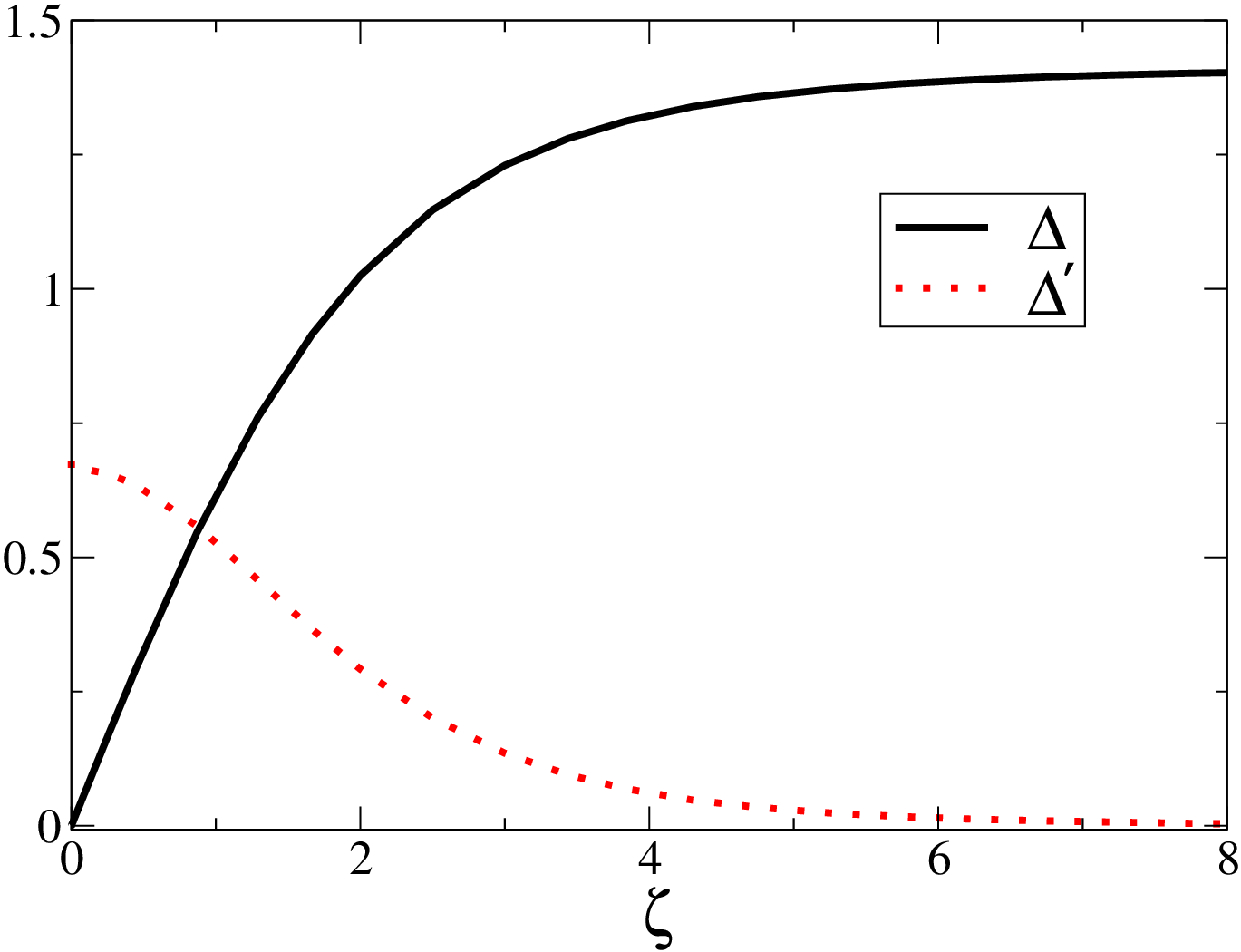}
\includegraphics[width=2.3in,angle=0]{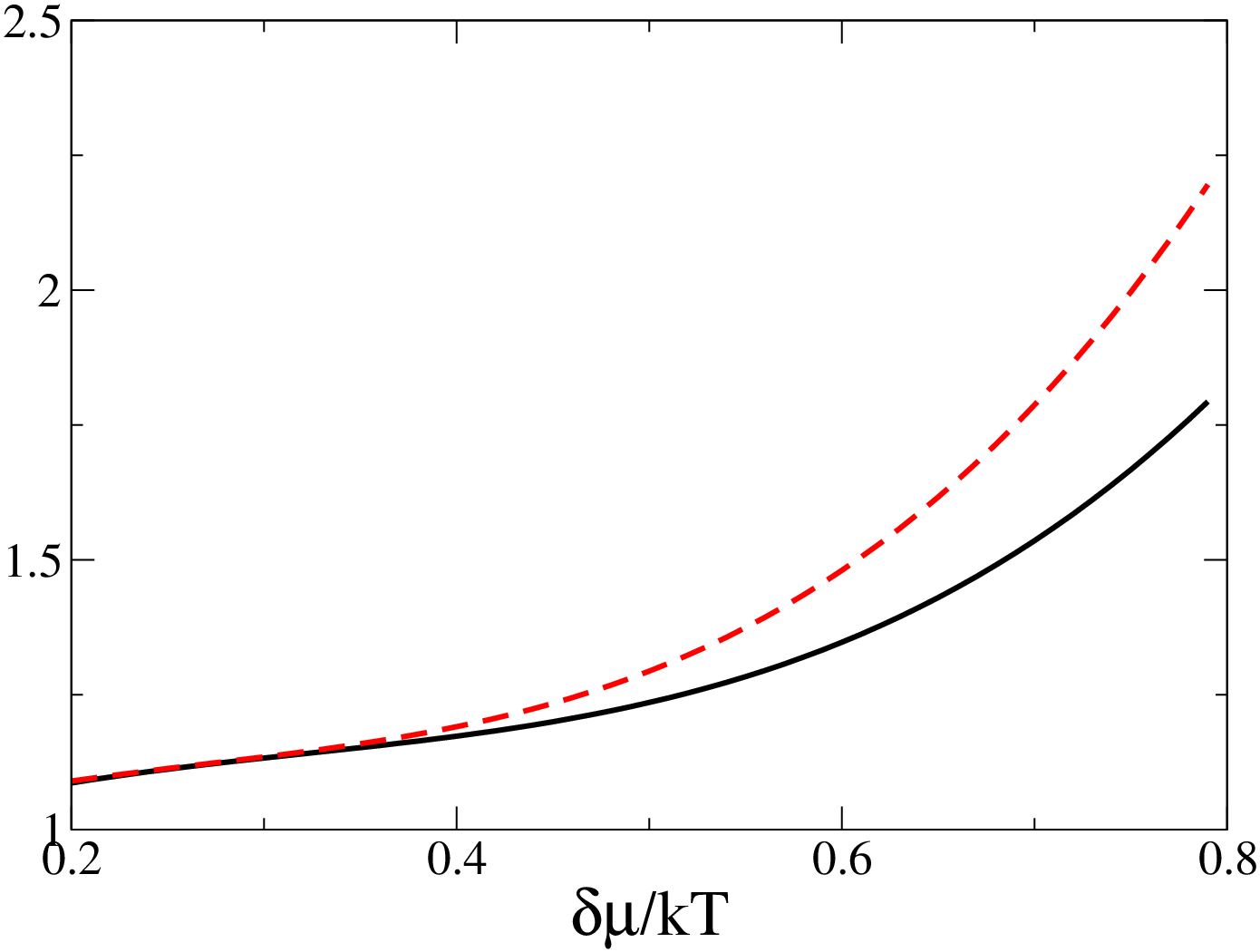}
\caption{(color online) Left panel: Numerical solution of the boundary value problem ODE, Eq.~(\ref{tdgl0}). The center of the vortex is at the origin of the axes coordinate and we report the plots of the condensate, full line,
and its derivative, dotted red line, as a function of the  distance from the center of the vortex. The condensate saturates at the boundary value,
$\Delta (r\rightarrow\infty)=1.4$. Parameters are $\mu =  -1$,
$\delta\mu=0.4$, $m=10$, $a_s=0.5$, where we are using units of $T$. 
Right panel: Normalized vortex size as a function of the  Fermi momenta mismatch. The vortex radius was extracted from the condensate configuration by two methods: based on the condensate (as it reaches
the value $\Delta=0.7  $), lower full curve,  and the condensate derivative (as it reaches $\Delta'=0.03 $), upper dashed red curve.
There is a transition to the gapless state around $\delta\mu=0.6$. At this point the slope of the upper curve increases.
The slope of the lower curve changes at this point too, but the curve is smoother.}  
\label{tdglfig}
\end{center}
\end{figure}

We considered the outer core radius, Eq. (\ref{r0}), which is obtained from an
expansion around the nontrivial vacuum state $\Delta\neq 0$. To further analyze
the vortex structure, we obtain the inner core radius which uses an expansion
around $\Delta=0$ state, i.e. Ginsburg-Landau expansion.  In \ref{ginsburg-landau} we
construct the Ginsburg-Landau functional to the fourth order, and derive the
equation  obeyed by  $\eta(r)$ in a vortex configuration using the Time
Dependent Ginzburg Landau equation (TDGL)~\cite{Melo},
\begin{equation}
\left(a+b\eta(r)^2-\frac{c}{2m}\nabla^2\right)\eta(r)=0
\label{tdgl}
\end{equation}
with the boundary conditions $\eta(r=0)=0$, $\eta(r\rightarrow\infty)=\eta_0$.
The center of the vortex is at $r=0$, and the expressions for the coefficients
$a, b, c$ are reported in
Eq.~(\ref{coef}).
 Notice that the TDGL equations for the vortex configuration are valid for $T\sim T_c$ where the gap is vanishing small. Introducing $\eta(r)={\rm e}^{i\phi}f(\zeta)\eta_0$, with  $\zeta=r\sqrt{2m\eta_0}$ a dimensionless variable,
we obtain the TDGL equation for the radial part of the condensate configuration  
\be
 \tilde{c}\left(\frac{1}{\zeta}\frac{d}{d\zeta}\left(\zeta\frac{df}{d\zeta}\right)
-\frac{f}{\zeta^2}\right)+\tilde{a}f-\tilde{b}f^3=0 
\label{tdgl0}
\ee
with boundary conditions $f(0)=0$ and  $f(\infty)=1$, and 
where the expression for the coefficients $\tilde{a}$, $\tilde{b}$, $\tilde{c}$ are reported in Eq.~(\ref{abc}).
At nonzero $T$, we solve numerically the TDGL equation for various values of
$\delta\mu$. In the left panel  of  Fig.~\ref{tdglfig}  we report the result
of the numerical solution of the condensate  and of its derivative for
$\delta\mu/T=0.4$.  In the right panel of Fig.~\ref{tdglfig}  we report the
value of the vortex radius as a function of $\delta\mu$.  The two curves
correspond to two different definitions of the vortex radius. based on a certain
value of either the condensate or it's derivative.
Although calculations
are done at nonzero $T$, one may assume that the same trend holds for vanishing temperatures.




\section{Conclusions}
In conclusion, our results show that as we move into the gapless regime, the
outer radius of the vortex increases sharply. This rise may be observed
in experiments done with cold atomic gases trapped in a magnetic trap.  
  
If the parameters of the trap, namely, the number of particles of the two
species, $N_1$ and $N_2$, and the scattering length, $a$, are tuned such that
there is a sufficiently wide region in position space where the atomic system is
in the gapless BEC phase, this dramatic effect can be seen. We leave the precise
determination of parameters $a$, $N_1$ and $N_2$ for future work, but it will
presumably require very flat traps to realize this phenomenon in a wide enough
region in the system to be observed cleanly.

The properties of vortices in the gapless region have been studied previously
in~\cite{stojanovic-2006,stojanovic-2008-323} who have concentrated on the
interaction between two vortices in this regime.    The vortex core structure in imbalanced superfluids
 has been studied in~\cite{Takahashi:2006} who have focused on the   occupation number of particles  that determine the
 ``visibility" of  vortices. The authors of \cite{iskin} used a Bogoliubov-de Gennes
 approach to solve for a vortex core state in fermion mixtures with unequal masses.
They found that the vortex core is mostly occupied by the light mass fermions and that the core density
of the heavy-mass fermions is highly depleted. We believe that their study points towards
the gapless phases, however their calculations are more involved. 
Our study provides motivation
to study a new observable, namely the size of the core of a vortex, in the
gapless phase.
 
Finally, we comment about the instability toward the formation of a 
non homogeneous phase. One can
see Fig.~\ref{contours} that in
the strong coupling regime the coefficient $B$ is always positive.
Indeed, along the curves corresponding to $\kappa=1$ and $\kappa=1.71$ in
Fig.~\ref{contours} the coefficient $B$ is positive and large. This
means that there is no instability toward a LOFF-like phase. This is
consistent with  the results of Ref.~\cite{Mannarelli:2006hr}, where
the LOFF phase was found to be favored in the weak coupling regime
only. Indeed from Fig.~\ref{contours} one can see that
in the weak coupling limit, the curves corresponding to $\kappa=0$ or
$\kappa=-0.5$ pass through  the region where $B$ is negative 
and this indicates that it is possible to
have a non homogeneous LOFF phase. But to really check the favorability of a 
LOFF-like phase in this region, in a small $\eta$ calculation, 
one should expand around the solution with $\Delta =0$ and not $\Delta \neq 0$. The 
reason being that the phase transition from (some) LOFF phases to the normal phase is
second order and one can study how fluctuations drive the system
from the homogeneous normal phase to a non homogeneous phase.
We leave such an analysis for future work.

\section{Acknowledgement}
The authors thank Andreas Schmitt and Sanjay Reddy for their valuable comments
on the manuscript. EG and RS thank Michael Forbes, Dam Son, Misha Stephanov,
Eugene Demler, Bertrand Halperin, Martin Zwierlein, Leonid Levitov
and Carlos Sa de Melo for discussions.
RS acknowledges several discussions with Sanjay Reddy. The work of MM has been
supported by the Ministerio de Educaci\'on y Ciencia (MEC) and CPAN under
grants FPA2007-66665 and 2009SGR502. RS is supported by LANS, LLC for the NNSA of the DOE under
contract $\#$DE-AC52-06NA25396. 

\appendix

\section{Relations between the Meissner mass, the Debye mass and the
coefficients of the effective action describing the phase
fluctuations} \label{appendix-relation}

 We show the equivalence between  the screening masses and the coefficients in
effective action for the Nambu-Goldstone mode. This equivalence can be
anticipated from gauge invariance if we gauge the global symmetry
associated with total number conservation~\cite{Son:1999cm}. The gauged action is
invariant under a local rotation in the phase of the fermion fields
$\psi_\beta$, with  $\beta=1,2$, accompanied by gauge transformation on the four vector
$(A_0,\bfA)$. (We apologize to the reader that we use the same Latin
character for the gauge fields as well as for the operator defined
in Eq.~(\ref{define A}). They can be easily distinguished because gauge fields
always appear with a subscript ($A_0$) or in bold font ($\bfA$).) The condensate
in Eq.~(\ref{condensate}) spontaneously breaks this gauge symmetry, and
therefore by the Anderson-Higgs mechanism, the gauge field components $\bf A$,
acquire a Meissner screening mass. The $A_0$ component of the gauge field is
instead Debye screened. (In the gapless regime, there is an additional
contribution to the Debye mass from the fermions in the blocking regions, that
we do not consider here~\cite{Gatto:2007ja}. Adding this contribution to the
pairing contribution that we calculate, gives the net Debye mass square for the
system.) In this Section we explicitly show that the screening masses can be
related to the coefficients that appear in the effective Lagrangian describing
the Nambu-Goldstone bosons.

 On gauging the quadratic part of the Nambu-Gorkov action in Eq.~(\ref{Seff1}) we obtain
\begin{equation}
\begin{split}
&{\cal{L}}=\bigl(\psi_1^*,\psi_2\bigr)\\
&
\Biggl(\begin{array}{cc}
{i\partial_{t}}+\frac{(\nabla-ig{\bf A})^2}{2m}+gA_0+\mu+\dm & -\Delta    \\
-\Delta  &  {i\partial_{t}}-\frac{(\nabla+ig{\bf A})^2}{2m}-gA_0-\mu+\dm
\end{array} \Biggr)
\bigl(\begin{array}{c}
 \psi_1\\
 \psi_2^*
\end{array}
\bigr)\\ 
&+(\dm\rightarrow -\dm) \label{gauged lagrangian}\,.
\end{split}
\end{equation}
 A gauge transformation is given by  $\psi_\beta\rightarrow \psi_\beta {\rm e}^{i\alpha(x)}$,
for fermions, leading to $\langle\psi_1\psi_2^*\rangle\propto \Delta(x)\rightarrow
\Delta{\rm e}^{2i\alpha(x)}$ for the condensate, and  by ${\bf A}\rightarrow {\bf
A}+\frac{1}{g}\nabla\alpha$, $A_0\rightarrow A_0+\frac{1}{g}\partial_t\alpha$
for the gauge field. Since the term with $\dm\rightarrow -\dm$ is common 
in all the following expressions, we will stop writing it explicitly from
Eq.~(\ref{quad_piece}) to Eq.~(\ref{s3}) and carry it implicitly, and only write it in the final
expression Eq.~(\ref{screening}).

First we establish the coefficients in the effective action of Nambu-Goldstone mode, in
the absence of any external gauge fields.
These are space-time dependent phase rotations of the $\Delta$-field which we parameterize
as $\Delta\rightarrow\Delta\exp(2i\alpha)$. The quadratic piece of the
Nambu-Gorkov fields has the form,
\begin{eqnarray}
&&\bigl(\psi_1^*,\psi_2\bigr)
\left(\begin{array}{cc}
{i\partial_{t}}+\frac{1}{2m}\nabla^2+\mu+\dm & -\Delta{\rm e}^{2i\alpha}    \\
-\Delta{\rm e}^{-2i\alpha}  &  {i\partial_{t}}-\frac{1}{2m}\nabla^2-\mu+\dm
\end{array} \right)
\bigl(\begin{array}{c}
 \psi_1\\
 \psi_2^*
\end{array}
\bigr)\nonumber\\
&=&\left(\psi_1^*,\psi_2\right)\left(\hat{O}+\hat{V}\right)
\bigl(\begin{array}{c}
 \psi_1\\
 \psi_2^*
\end{array}
\bigr)\label{quad_piece}\,
\end{eqnarray}
where $\hat{O}$ and $\hat{V}$ are given by Eq. (\ref{define O}).
To the relevant order this phase shifts $\Delta$ to $\Delta+\eta$, with
fluctuations $\eta$ being given by $\eta=\Delta(2i\alpha)$.
The effective action for $\alpha$ is obtained by integrating out the fermions
and can be written directly from Eq.~(\ref{Action}) by substituting this value
of $\eta$. We obtain,
\begin{equation}
{\cal{S}}^{(2)}_{\rm
Goldstone}=-\Delta^2\left(\frac{T}{V}\right)^2\sum_{k,p}\alpha(k)\alpha(-k)\left\{
\frac{k_0^2-(\xi({\bf p}+{\bf k})-\xi({\bf p}))^2}{D(p)D(p+k)}\right\}
\label{action}\;,
\end{equation}
where $k_0=ik_4$.

Next we remove the phase from the condensate by redefining the phases of the
fermionic fields. This will give rise to non-zero values of the gauge fields and
we evaluate  the screening masses of these gauge fields. Redefining
$\psi_\beta=\tilde{\psi}_\beta\exp{(i\alpha)}$ and acting with a derivative operator, the quadratic
part of the action  can be written as
\begin{eqnarray}
&&\bigl(\tilde{\psi_1}^*{\rm e}^{-i\alpha},\tilde{\psi}_2{\rm e}^{i\alpha}\bigr)
\left(\begin{array}{cc}
{i\partial_{t}}+\frac{1}{2m}\nabla^2+\mu+\dm & -\Delta{\rm e}^{2i\alpha}    \\
-\Delta{\rm e}^{-2i\alpha}  &  {i\partial_{t}}-\frac{1}{2m}\nabla^2-\mu+\dm
\end{array} \right)
\bigl(\begin{array}{c}
 \tilde{\psi}_1{\rm e}^{i\alpha}\\
 \tilde{\psi}_2^*{\rm e}^{-i\alpha}
\end{array}
\bigr)\nonumber\\
&=&\bigl(\tilde{\psi}_1^*,\tilde{\psi}_2\bigr)\left(\hat{O}+\tilde{V}\right)
\bigl(\begin{array}{c}
 \tilde{\psi}_1\\
 \tilde{\psi}_2^*
\end{array}
\bigr)
\end{eqnarray}
where $\hat{O}$ is given by Eq. (\ref{define O}), and $\tilde{V}$ is given by
\begin{equation}
\tilde{V} = \left(
\begin{array}{cc}
-\partial_t\alpha+\frac{(\nabla\cdot i\nabla\alpha+i\nabla\alpha\cdot \nabla)}{2m}-\frac{(\nabla\alpha)^2}{2m} & 0    \\
0 & \partial_t\alpha+\frac{(\nabla\cdot i\nabla\alpha+i\nabla\alpha\cdot \nabla)}{2m}+\frac{(\nabla\alpha)^2}{2m}
\end{array} \right)\;,
\end{equation}
which includes first and second order terms in $\alpha$. This Lagrangian is
exactly of the form given by Eq.~(\ref{gauged lagrangian}) with
$g(A_0,\bfA)=(-\partial_t\alpha,-{\bf{\nabla}}\alpha)$ and hence the gauge boson
masses can be read from the Lagrangian describing the $\alpha$ fields.

To show that the masses we obtain this way are the same as the
coefficients obtained by treating $\alpha$ as the Nambu-Goldstone field
(Eq.~(\ref{action})) we explicitly calculate the second order
correction to the action using Eq.~(\ref{expansion}). We will
analyze separately the quadratic  term that contain only spatial
derivatives of $\alpha$,  the term that contain only time
derivatives of $\alpha$ and the mixed term.

Consider the quadratic spatial component, $({\bf \nabla}\alpha)^2$.  Both linear
and quadratic  terms in $\tilde{V}$ contribute to this  part of the
effective action. Indeed from the expansion of the action  we obtain
\begin{equation}
\begin{split}
-2{{\cal{S}}_{\rm{spatial}}^{(2)}}
=&\tr(\hat{O}^{-1}\tilde{V})-\frac{1}{2}\tr(\hat{O}^{-1}\tilde{V}\hat{O}^{-1}\tilde{V})|_{\partial_t\alpha=0}\label{spacial}\\
=& \left(\frac{T}{V}\right)^2\sum_{k,p}\alpha(k)\alpha(-k)\Bigl\{
-\frac{k^2}{2m}\frac{\tilde{A}(p)-A(p)}{D(p)}\\
&\phantom{++++}-\frac{1}{2}\frac{((2{\bf p}+{\bf k})\cdot{\bf k})^2}{(2m)^2}
\frac{\tilde{A}(p)\tilde{A}(p+k)+A(p)A(p+k)+2\Delta^2}{D(p)D(p+k)}
\Bigr\}\;.
\end{split}
\end{equation}
Note that differential operator in $\nabla\alpha$ acts on external in- and out-going legs, i.e. produces
$i{\bf k}$ and $-i{\bf k}$, while the other $\nabla$ acts inside the loop, producing $i{\bf p}$ and
$i({\bf p}+\bf {k})$. Using the definitions of $A(p)$ and $\tilde{A}(p)$
(Eq.~(\ref{define A}))
the term proportional to $\tilde{A}(p)-A(p)$ on the right hand side of  Eq. (\ref{spacial}) simplifies to
\begin{equation}
\Bigl(\frac{T}{V}\Bigr)^2\sum_{k,p}\alpha(k)\alpha(-k)\Bigl\{
-\frac{k^2}{m}\frac{\xi({\bf p})}{D(p)}\Bigr\} \;.
\end{equation}
The second term on the right hand side of Eq. (\ref{spacial})
can be simplified by noticing that \begin{equation}
\frac{(2{\bf p}+{\bf k})\cdot{\bf k}}{2m}=\xi({\bf p}+{\bf k})-\xi({\bf p})
=A({ p})-A({ p + k})+k_0=\tilde{A}({ p + k})-\tilde{A}({ p})-k_0\label{simplify}\;.
\end{equation}
Then the  second term on the rhs  of Eq.~(\ref{spacial}) is given by
\begin{eqnarray}
&&\Bigl(\frac{T}{V}\Bigr)^2\sum_{k,p}\alpha(k)\alpha(-k)\Bigl\{
-\frac{1}{2}(\xi({\bf p})-\xi({\bf p}+{\bf k}))^2\;
\frac{\tilde{A}(p)\tilde{A}(p+k)+A(p)A(p+k)+2\Delta^2}{D(p)D(p+k)}
\Bigr\}\nonumber\\
&=&\Bigl(\frac{T}{V}\Bigr)^2\sum_{k,p}\alpha(k)\alpha(-k)\Bigl\{
-\Delta^2\frac{2(\xi({\bf p})-\xi({\bf p}+{\bf k}))^2}{D(p)D(p+k)}\Bigr.\nonumber\\ &&
-\Bigl.\frac{1}{2}k_0(\xi({\bf p})-\xi({\bf p}+{\bf k}))\;
\frac{\tilde{A}(p)\tilde{A}(p+k)-A(p)A(p+k))}{D(p)D(p+k)}
 \Bigr\}
-\Bigl(\frac{T}{V}\Bigr)^2 r\,,
\end{eqnarray}
where
\begin{eqnarray}
r & =&
\sum_{k,p}\alpha(k)\alpha(-k)\Bigl\{
\frac{(\xi({\bf p})-\xi({\bf p}+{\bf k}))}{2D(p)}(\tilde{A}(p)-A(p))
-\frac{(\xi({\bf p})-\xi({\bf p}+{\bf k}))}{2D(p+k)}(\tilde{A}(p+k)-A(p+k)) \Bigr\}  \nonumber\\
&=&\sum_{k,p}\alpha(k)\alpha(-k)\Bigl\{
\frac{\xi({\bf p})}{D(p)}(2\xi({\bf p})-\xi({\bf p}-{\bf k})-\xi({\bf p}+{\bf k})) \Bigr\} 
=\sum_{k,p}\Bigl\{
-\frac{\xi({\bf p})}{D(p)}\frac{k^2}{m}\Bigr\}  \nonumber \,.
\end{eqnarray}
Here we have used Eq.~(\ref{simplify}) to rewrite the combination $\xi({\bf p})-\xi({\bf p}+{\bf k})$
and to extract extra power of $D(p)=\tilde{A}(p)A(p)-\Delta^2$ and $D(p+k)$ in the numerator.
Combining all the terms together, we get for the spatial component of the  effective action
\begin{equation}
\begin{split}
&{\cal{S}}^{(2)}_{\rm
spatial}=\Bigl(\frac{T}{V}\Bigr)^2\sum_{k,p}\alpha(k)\alpha(-k)\Bigl\{
\Delta^2\frac{(\xi({\bf p})-\xi({\bf p}+{\bf k}))^2}{D(p)D(p+k)}\Bigr.\\
&+\Bigl.\frac{1}{4}k_0(\xi({\bf p})-\xi({\bf p}+{\bf k}))\;
\frac{\tilde{A}(p)\tilde{A}(p+k)-A(p)A(p+k)}{D(p)D(p+k)}
\Bigr\}\label{s1}\;.
\end{split}
\end{equation}

Now we turn to the quadratic term containing only temporal derivatives, i.e.
$(\partial_t\alpha)^2$. Only terms quadratic in   $\tilde{V}$  contribute to
this  part of the effective action (Eq. (\ref{expansion})),
\begin{equation}
\begin{split}
&-2{\cal{S}}^{(2)}_{\rm
temporal}=-\frac{1}{2}\tr(\hat{O}^{-1}\tilde{V}\hat{O}^{-1}\tilde{V})|_{{\bf{\nabla}}\alpha=0}\\
&=\Bigl(\frac{T}{V}\Bigr)^2\sum_{k,p}\alpha(k)\alpha(-k)\Bigl\{
-\frac{k_0^2}{2}
\frac{\tilde{A}(p)\tilde{A}(p+k)+A(p)A(p+k)-2\Delta^2}{D(p)D(p+k)}
\Bigr\}\label{temporal}\;.
\end{split}
\end{equation}
We can simplify this expression, noticing that one can write
\begin{equation}
k_0=\tilde{A}(p)-\tilde{A}(p+k)-(\xi({\bf p})-\xi({\bf p}+{\bf k}))
=A(p)-A(p+k)+(\xi({\bf p})-\xi({\bf p}+{\bf k}))
\end{equation}
whereupon the right hand side of Eq. (\ref{temporal}) can be rewritten as
\begin{equation}
\begin{split}
\Bigl(\frac{T}{V}\Bigr)^2&\sum_{k,p}\alpha(k)\alpha(-k)\Bigl\{
\Delta^2\frac{2k_0^2}{D(p)D(p+k)}\\
&-\frac{1}{2}k_0(\xi({\bf p})-\xi({\bf p}+{\bf k}))\;
\frac{\tilde{A}(p)\tilde{A}(p+k)-A(p)A(p+k)}{D(p)D(p+k)}\Bigr\}
-\Bigl(\frac{T}{V}\Bigr)^2\Bigl\{s\Bigr\}\;,
\end{split}
\end{equation}
where,
\begin{equation}
s=\sum_{k,p}\alpha(k)\alpha(-k)
\frac{1}{2}k_0\Bigl(\frac{\tilde{A}(p+k)+A(p+k)}{D(p+k)}-\frac{\tilde{A}(p)+A(p)}{D(p)}\Bigr)=0 \,.
\end{equation}
Then, the term of the effective action containing the temporal derivatives of $\alpha$ turns out to be given by
\begin{equation}
\begin{split}
&{\cal{S}}^{(2)}_{\rm temporal}=\Bigl(\frac{T}{V}\Bigr)^2\sum_{k,p}\alpha(k)\alpha(-k)\Bigl\{
-\Delta^2\frac{k_0^2}{D(p)D(p+k)}\Bigr.\\
&+\Bigl.\frac{1}{4}k_0(\xi({\bf p})-\xi({\bf p}+{\bf k}))\;
\frac{\tilde{A}(p)\tilde{A}(p+k)-A(p)A(p+k)}{D(p)D(p+k)}
\Bigr\}\label{s2}\;.
\end{split}
\end{equation}

Finally we consider the mixed component, $(\nabla\alpha)(\partial_t\alpha)$. The only
contribution to this part of the action comes from the quadratic term in $\tilde{V}$
(Eq. (\ref{expansion})). Therefore,
\begin{equation}
\begin{split}
&-2{\cal{S}}^{(2)}_{\rm
mixed}=-\frac{1}{2}\tr(\hat{O}^{-1}\tilde{V}\hat{O}^{-1}\tilde{V})|_{{\rm{mixed}}}\\
&= \Bigl(\frac{T}{V}\Bigr)^2\sum_{k,p}\alpha(k)\alpha(-k)\Bigl\{
k_0\frac{(2{\bf p}+{\bf k})\cdot{\bf k}}{2m}
\frac{\tilde{A}(p)\tilde{A}(p+k)-A(p)A(p+k)}{D(p)D(p+k)}
\Bigr\}\label{mixed}\;.
\end{split}
\end{equation}
We use Eq. (\ref{simplify}) to simplify the mixed term, and we obtain the following contribution to the effective action
\begin{equation}
{\cal{S}}^{(2)}_{\rm
mixed}=\Bigl(\frac{T}{V}\Bigr)^2\sum_{k,p}\alpha(k)\alpha(-k)\Bigl\{
-\frac{1}{2}k_0(\xi({\bf p})-\xi({\bf p}+{\bf k}))\;
\frac{\tilde{A}(p)\tilde{A}(p+k)-A(p)A(p+k)}{D(p)D(p+k)}
\Bigr\}\label{s3}\;.
\end{equation}

The sum of the  three  terms, Eq. (\ref{s1}),(\ref{s2}) and (\ref{s3}), gives the effective action for the gauge field
\begin{equation}
\begin{split}
{\cal{S}}^{(2)}=&{\cal{S}}^{(2)}_{\rm spatial}+{\cal{S}}^{(2)}_{\rm temporal}+{\cal{S}}^{(2)}_{\rm mixed}=
-\Delta^2\Bigl(\frac{T}{V}\Bigr)^2\sum_{k,p}\alpha(k)\alpha(-k)\Bigl\{
\frac{k_0^2-(\xi({\bf p})-\xi({\bf p}+{\bf k}))^2}{D(p)D(p+k)}
\Bigr\}\\
&+(\dm\rightarrow -\dm)\label{screening}\;.
\end{split}
\end{equation}
We recognize that by putting $k_0=0$ in ${\cal{S}}^{(2)}$ we reproduce the Meissner mass, and by putting
${\bf k}=0$, i.e. $\xi({\bf p})-\xi({\bf p}+{\bf k})=0$, we obtain the Debye mass.
Comparing Eq. (\ref{action}) and Eq. (\ref{screening}), we see that coefficients in the effective action for the Nambu-Goldstone mode
coincide with the corresponding screening masses as we set out to show.

\section{Coefficients of the low energy Lagrangian}
\label{appendix-coefficients}
The sum over the  Matsubara frequencies can be done analytically noticing   that
if  $f(x)$ is a function with  no poles then one has 
\begin{equation}
T\sum_{n= -\infty}^{\infty} \frac{f(i \omega_n)}{i\omega_n + {\cal{E}}} = \ha
\tanh[{\cal E}/(2T)]\, f(-{\cal{E}})\label{p4 integration formula}\;,
\end{equation}
for $\omega_n = (2n + 1 ) \pi T$.
Upon substituting  this result in  Eq.~(\ref{gap equation}) one obtains  the
usual form of the gap equation at non vanishing  temperatures:
\begin{equation}\label{gapT}
\frac{1}{\lambda} = \frac{1}{2V} \sum_{\vp} \frac{1}{\epsilon(\vp)} g(\epsilon(\vp))\;,
\end{equation}
where for infinite volume, the sum over $\bf p$ can be replaced by
an integral over three-momentum $\bf p$ and where  we have defined,
\be
g(\epsilon)= \ha\Bigl(\tanh\bigl[\frac{\dmu+\epsilon}{2T}\bigl] +
\tanh\bigl[\frac{-\dmu+\epsilon}{2T}\bigl] \Bigr)\nonumber =  n_f(-\dmu-\epsilon) - n_f(-\dmu+\epsilon)\;,
\ee
with $n_f$  the Fermi-Dirac distribution function.

 After  evaluating the Matsubara $p_4$ sums using Eq.~(\ref{p4 integration
formula}), we get the following expressions,
\begin{eqnarray}
I_1(k)&=&\frac{\Delta^2}{16}\frac{1}{V}\sum_{\vp}\frac{1}{\epsilon_1\epsilon}
\Bigl\{ \bigl(g(\epsilon_1)-g(\epsilon)\bigr)
\Bigl(\frac{1}{k_0+\epsilon_1-\epsilon}-\frac{1}{k_0-\epsilon_1+\epsilon}\Bigr)\nonumber\\
&+& \bigl(g(\epsilon_1)+g(\epsilon)\bigr)
\Bigl(\frac{1}{k_0-\epsilon_1-\epsilon}-\frac{1}{k_0+\epsilon_1+\epsilon}\Bigr)\Bigr\}\nonumber\\
I_2(k)&=&\frac{-1}{16} \frac{1}{V}\sum_{\vp}
\frac{k_0^2-(\xi-\xi_1)^2}{\epsilon_1\epsilon}
 \Bigl\{\bigl(g(\epsilon_1)-g(\epsilon)\bigr)
\Bigl(\frac{1}{k_0+\epsilon_1-\epsilon}-\frac{1}{k_0-\epsilon_1+\epsilon}\Bigr)\nonumber\\
&+&\bigl(g(\epsilon_1)+g(\epsilon)\bigr)
\Bigl(\frac{1}{k_0-\epsilon_1-\epsilon}-\frac{1}{k_0+\epsilon_1+\epsilon}\Bigr)\Bigr\}\nonumber\\
I_3(k)&=& \frac{1}{8} \frac{1}{V}\sum_{\vp}
\frac{1}{\epsilon_1\epsilon}
\Bigl\{\bigl(g(\epsilon_1)-g(\epsilon)\bigr)\bigl(\epsilon\xi_1-\epsilon_1\xi\bigr)
\Bigl(\frac{1}{k_0+\epsilon_1-\epsilon}+\frac{1}{k_0-\epsilon_1+\epsilon}\Bigr)\nonumber\\
&+&
\bigl(g(\epsilon_1)+g(\epsilon)\bigr)\bigl(\epsilon\xi_1+\epsilon_1\xi\bigr)
\Bigl(\frac{1}{k_0-\epsilon_1-\epsilon}+\frac{1}{k_0+\epsilon_1+\epsilon}\Bigr)\Bigr\}\label{I123}\;,
\end{eqnarray}
where the sum is an integral over the three-momentum $\bf p$ and we have indicated with $\epsilon_1$ and $\xi_1$ the quantities $\epsilon(\vp+\vk)$ and
$\xi(\vp+\vk)$ respectively, with $\epsilon$ and $\xi$ the quantities
$\epsilon(\vp)$ and $\xi(\vp)$ respectively, and where $k_0=ik_{4}$.

The expressions of the coefficients  $A$, $B$, $C$, $D$, $E$ and $F$ in
Eq.(\ref{define ABCDEF}) are given by
\begin{eqnarray}
A&=&\frac{1}{8}\frac{1}{V}\sum_\vp  
\frac{g}{\epsilon^3} \nonumber\\
B&=& \frac{1}{8}\frac{1}{V}\sum_\vp  \Bigl\{
g
\frac{\vp^2}{m^2\epsilon^3}
-g'
\frac{\vp^2}{m^2\epsilon^2} \Bigr\}
=\frac{1}{4}\frac{1}{V}\sum_\vp \Bigl\{
g\frac{\xi+\mu}{m\epsilon^3}-g'\frac{\xi+\mu}{m\epsilon^2}
\Bigr\}\nonumber\\
C&=&\Delta^2\frac{1}{2}\frac{1}{V}\sum_\vp  \Bigl\{
\frac{g}{\epsilon^3}-\frac{g'}{\epsilon^2}\Bigr\}\nonumber\\
D&=&\frac{1}{8}\frac{1}{V}\sum_\vp  
g\frac{\xi^2}{\epsilon^5}\nonumber\\
E&=&\frac{1}{8}\frac{1}{V}\sum_\vp\Bigl\{
g\Bigl[-\frac{9\xi\Delta^2}{\epsilon^5m}+\frac{\vp^2(\xi^4+9\xi^2\Delta^2-2\Delta^4)}{\epsilon^7m^2}
\Bigr]\nonumber\\
&-&g'\Bigl[-\frac{9\xi\Delta^2}{\epsilon^4m}+\frac{\vp^2(\xi^4+9\xi^2\Delta^2-2\Delta^4)}{\epsilon^6m^2}
\Bigr]\nonumber\\
&-&g''\Bigl[\frac{3\xi\Delta^2}{\epsilon^3m}+\frac{\vp^2(-3\xi^2\Delta^2+\Delta^4)}{\epsilon^5m^2}
\Bigr]\nonumber\\
&-&g'''\frac{2\vp^2\xi^2\Delta^2}{3\epsilon^4m^2}
\Bigr\}\nonumber\\
&=&\frac{-3}{8}\frac{1}{V}\sum_\vp  \Bigl\{
g\Bigl[
\frac{\xi}{3m}\Bigl(\frac{20\Delta^4}{\epsilon^7}-\frac{5\Delta^2}{\epsilon^5}-\frac{2}{\epsilon^3}\Bigr)
+ \frac{2\mu}{3m}\Bigl(\frac{10\Delta^4}{\epsilon^7}-\frac{7\Delta^2}{\epsilon^5}-\frac{1}{\epsilon^3}\Bigr)\Bigr]\nonumber\\
&-&g'\Bigl[
\frac{\xi}{3m}\Bigl(\frac{20\Delta^4}{\epsilon^6}-\frac{5\Delta^2}{\epsilon^4}-\frac{2}{\epsilon^2}\Bigr)
+ \frac{2\mu}{3m}\Bigl(\frac{10\Delta^4}{\epsilon^6}-\frac{7\Delta^2}{\epsilon^4}-\frac{1}{\epsilon^2}\Bigr)
\Bigr] \nonumber\\
&+&g''\Bigl[
\frac{\xi}{3m}\Bigl(\frac{8\Delta^4}{\epsilon^5}-\frac{3\Delta^2}{\epsilon^3}\Bigr)
+ \frac{2\mu}{3m}\Bigl(\frac{4\Delta^4}{\epsilon^5}-\frac{3\Delta^2}{\epsilon^3}\Bigr)
\Bigr] \nonumber\\
&+&g'''\Bigl[
\frac{4\xi}{9m}\Bigl(\frac{-\Delta^4}{\epsilon^4}+\frac{\Delta^2}{\epsilon^2}\Bigr)
+\frac{4\mu}{9m}\Bigl(\frac{-\Delta^4}{\epsilon^4}+\frac{\Delta^2}{\epsilon^2}\Bigr)
\Bigr]\Bigr\}\nonumber\\
F&=&\frac{1}{4}\frac{1}{V}\sum_\vp  
g\frac{\xi}{\epsilon^3}\label{expressionsABCDEF}\;,
\end{eqnarray}
where $g'$ refers to the differentiation of $g(\epsilon)$ with respect to
$\epsilon$. 

To evaluate the integrals we use the following relations. For any function
$f(\epsilon,\xi)$ we have,
\begin{eqnarray}
\threeier{\magp} gf &=& \frac{1}{2\pi^2}\Biggl[
\int_{0}^{\magp_-} d\magp \magp^2f +
\int_{\magp_+}^{\infty} d\magp \magp^2f \Biggr]\nonumber\\
\threeier{\magp} \magp^2gf &=& \frac{1}{2\pi^2}\Biggl[
\int_{0}^{\magp_-} d\magp \magp^4f +
\int_{\magp_+}^{\infty} d\magp \magp^4f \Biggr]\nonumber\\
\threeier{\magp} g'f &=&
\gapless\frac{1}{2\pi^2}\Biggl[\frac{m\delta\mu}{\sqrt{\dm^2-\Delta^2}}\Bigl(\magp f\Big|_{\magp_-}
+\magp f\Big|_{\magp_+}\Bigr)\Biggr]\nonumber\\
\threeier{\magp} \magp^2g'f &=&
\gapless\frac{1}{2\pi^2}\Biggl[\frac{m\delta\mu}{\sqrt{\dm^2-\Delta^2}}\Bigl(\magp^3f\Big|_{\magp_-}
+\magp^3f\Big|_{\magp_+}\Bigr)\Biggr]\nonumber\\
\threeier{\magp} g''f &=&
\gapless\frac{-1}{2\pi^2} \Biggl[\frac{m\delta\mu}{\sqrt{\dm^2-\Delta^2}}
\Bigl(\magp\frac{d}{d\xi}\Bigl(\frac{f\epsilon}{\xi}\Bigr)\Big|_{\magp_-}
+\magp\frac{d}{d\xi}\Bigl(\frac{f\epsilon}{\xi}\Bigr)\Big|_{\magp_+}\Bigr)\nonumber\\
&+&\frac{m^2\delta\mu^2}{(\dm^2-\Delta^2)}\Bigl(-\frac{f}{\magp}\Big|_{\magp_-} +
\frac{f}{\magp}\Big|_{\magp_+}\Bigr)\Biggr]\nonumber\\
\threeier{\magp} g'''f &=&
\gapless\frac{1}{2\pi^2} \Biggl[\frac{-\Delta^2m}{(\dm^2-\Delta^2)^{(3/2)}}
\Bigl({\magp_-}\frac{d}{d\xi}\Bigl(\frac{f\epsilon}{\xi}\Bigr)\Big|_{\magp_-}
+{\magp_+}\frac{d}{d\xi}\Bigl(\frac{f\epsilon}{\xi}\Bigr)\Big|_{\magp_+}\Bigr)\nonumber\\
&+&
\frac{-\Delta^2m^2\dm}{(\dm^2-\Delta^2)^{2}}
\Bigl(-\frac{f}{\magp_-}\Big|_{\magp_-}
+\frac{f}{\magp_+}\Big|_{\magp_+}\Bigr)\nonumber\\
&+&\frac{m\delta\mu^2}{({\dm^2-\Delta^2})}
\Bigl(-{\magp_-}\frac{d^2}{d\xi^2}\Bigl(\frac{f\epsilon}{\xi}\Bigr)\Big|_{\magp_-}
+{\magp_+}\frac{d^2}{d\xi^2}\Bigl(\frac{f\epsilon}{\xi}\Bigr)\Big|_{\magp_+}\Bigr)\nonumber\\
&+&\frac{2m^2\delta\mu^2}{({\dm^2-\Delta^2})}
\Bigl(-\frac{1}{\magp_-}\frac{d}{d\xi}\Bigl(\frac{f\epsilon}{\xi}\Bigr)\Big|_{\magp_-}
+\frac{1}{\magp_+}\frac{d}{d\xi}\Bigl(\frac{f\epsilon}{\xi}\Bigr)\Big|_{\magp_+}\Bigr)\nonumber\\
&+&\frac{m^3\delta\mu^3}{(\dm^2-\Delta^2)^{(3/2)}}\Bigl(-\frac{f}{(\magp_-)^3}\Big|_{\magp_-} -
\frac{f}{(\magp_+)^3}\Big|_{\magp_+}\Bigr)\Biggr]\;,~\label{general integrals}
\end{eqnarray}
where $\magp_{\pm}=\theta(\dmu-\Delta)\theta(\mu\pm\sqrt{\delta\mu^2-\Delta^2})\sqrt{2m(\mu\pm\sqrt{\delta\mu^2-\Delta^2})}$, $\epsilon({\magp_+})=\epsilon({\magp_-})=\dm$ and
$\xi({\magp_+})=-\xi({\magp_-})=\theta(\dmu-\Delta)\sqrt{\delta\mu^2-\Delta^2}$.
In Eq.~(\ref{general integrals}), all algebraic terms featuring $\magp_{\pm}$
appear with a corresponding product of $\theta$ functions,
$\theta(\dmu-\Delta)\theta(\mu\pm\sqrt{\delta\mu^2-\Delta^2})$, which we have
omitted for clarity. Whenever $\magp_{\pm}$ appear as limits of the integrals,
we can simply use the definitions of $\magp_{\pm}$ given above to obtain the correct answer.

We analyze the values of the coefficients $A,B,C,D,E,F$ at $T=0$. This can be done by taking the limit $T\rightarrow 0$ in
Eq.~(\ref{expressionsABCDEF}). We obtain the following expressions:
\begin{eqnarray}
A&=&\frac{1}{16\pi^2} \Biggl[
\int_{0}^{\magp_-} d\magp  \frac{\magp^2}{\epsilon^3} +
\int_{\magp_+}^{\infty} d\magp  \frac{\magp^2}{\epsilon^3} \Biggr]
\label{A final}\nonumber\\
B&=& \frac{1}{16m^2\pi^2} \Biggl[
\int_{0}^{\magp_-} d\magp  \frac{\magp^4}{\epsilon^3} +
\int_{\magp_+}^{\infty} d\magp  \frac{\magp^4}{\epsilon^3}
-\frac{m}{\delta\mu \sqrt{\dm^2-\Delta^2}}((\magp_-)^3
+(p_+)^3)
 \Biggr]\label{B final}\nonumber\\
C&=&\frac{\Delta^2}{4\pi^2} \Biggl[
\int_{0}^{\magp_-} d\magp  \frac{\magp^2}{\epsilon^3} +
\int_{\magp_+}^{\infty} d\magp  \frac{\magp^2}{\epsilon^3}-
 \frac{m}{\delta\mu\sqrt{\dm^2-\Delta^2}}(\magp_-
+p_+)   \Biggr]\label{C final}\nonumber\\
D&=&\frac{1}{16\pi^2} \Biggl[
\int_{0}^{\magp_-} d\magp  \frac{\magp^2\xi^2}{\epsilon^5} +
\int_{\magp_+}^{\infty} d\magp  \frac{\magp^2\xi^2}{\epsilon^5} \Biggr]\label{D final}\;,
\end{eqnarray}

The equation for $E$ is more complicated and hence we do not give the detailed  final expression which however can be obtained from the following equation:
\begin{eqnarray}
E&=&\frac{-3}{8m}\threeier{\magp}  \Bigl\{
g  R 
+g'  S 
+g'' T 
+g''' U \Bigr\}\nonumber \\
&=&\frac{-3}{16m\pi^2}  \Bigl\{
 \int_0^{\magp_-} d\magp \magp^2R + \int_0^{\magp_+} d\magp \magp^2R \nonumber\\
&+&\gapless\frac{m\dm}{\sqrt{\dm^2-\Delta^2}}\Bigl( {\magp_-}S\Big |_{p_-}
+{\magp_+}S\Big|_{\magp_+} \Bigr)\nonumber \\
&-&\gapless\Bigl[ \frac{m\delta\mu}{\sqrt{\dm^2-\Delta^2}}
\bigl[\magp_-\frac{d}{d\xi}\Bigl(\frac{T\epsilon}{\xi}\Bigr)\Big|_{\magp_-}
+\magp_+\frac{d}{d\xi}\Bigl(\frac{T\epsilon}{\xi}\Bigr)\Big|_{\magp_+}\bigr]\nonumber\\
&&\phantom{blah}+\frac{m^2\delta\mu^2}{(\dm^2-\Delta^2)}\bigl[-\frac{T}{\magp_-}\Big|_{\magp_-} +
\frac{T}{\magp_+}\Big|_{\magp_+}\bigr]\Bigr] \nonumber\\
&+&\gapless\Bigl[ \frac{-\Delta^2m}{(\dm^2-\Delta^2)^{(3/2)}}
\bigl[{\magp_-}\frac{d}{d\xi}\Bigl(\frac{U\epsilon}{\xi}\Bigr)\Big|_{\magp_-}
+{\magp_+}\frac{d}{d\xi}\Bigl(\frac{U\epsilon}{\xi}\Bigr)\Big|_{\magp_+}\bigr]\nonumber\\
&&\phantom{blah}+
\frac{-\Delta^2m^2\dm}{(\dm^2-\Delta^2)^{2}}
\bigl[-\frac{U}{\magp_-}\Big|_{\magp_-}
+\frac{U}{\magp_+}\Big|_{\magp_+}\bigr]\nonumber\\
&&\phantom{blah}+\frac{m\delta\mu^2}{({\dm^2-\Delta^2})}
\bigl[-{\magp_-}\frac{d^2}{d\xi^2}\Bigl(\frac{U\epsilon}{\xi}\Bigr)\Big|_{\magp_-}
+{\magp_+}\frac{d^2}{d\xi^2}\Bigl(\frac{U\epsilon}{\xi}\Bigr)\Big|_{\magp_+}\bigr]\nonumber\\
&&\phantom{blah}+\frac{2m^2\delta\mu^2}{({\dm^2-\Delta^2})}
\bigl[-\frac{1}{\magp_-}\frac{d}{d\xi}\Bigl(\frac{U\epsilon}{\xi}\Bigr)\Big|_{\magp_-}
+\frac{1}{\magp_+}\frac{d}{d\xi}\Bigl(\frac{U\epsilon}{\xi}\Bigr)\Big|_{\magp_+}\bigr]\nonumber\\
&&\phantom{blah}+\frac{m^3\delta\mu^3}{(\dm^2-\Delta^2)^{(3/2)}}\Bigl(-\frac{U}{(\magp_-)^3}\Big|_{\magp_-} -
\frac{U}{(\magp_+)^3}\Big|_{\magp_+}\Bigr) \Bigr]\Bigr\}
\end{eqnarray}
where,
\begin{eqnarray}
R&=& \frac{\xi}{3}\Bigl(\frac{20\Delta^4}{\epsilon^7}-\frac{5\Delta^2}{\epsilon^5}-\frac{2}{\epsilon^3}\Bigr)
+
\frac{2\mu}{3}\Bigl(\frac{10\Delta^4}{\epsilon^7}-\frac{7\Delta^2}{\epsilon^5}-\frac{1}{\epsilon^3}\Bigr)\nonumber\\
S&=&-\frac{\xi}{3}\Bigl(\frac{20\Delta^4}{\epsilon^6}-\frac{5\Delta^2}{\epsilon^4}-\frac{2}{\epsilon^2}\Bigr)
-
\frac{2\mu}{3}\Bigl(\frac{10\Delta^4}{\epsilon^6}-\frac{7\Delta^2}{\epsilon^4}-\frac{1}{\epsilon^2}\Bigr)\nonumber\\
T&=& \frac{\xi}{3}\Bigl(\frac{8\Delta^4}{\epsilon^5}-\frac{3\Delta^2}{\epsilon^3}\Bigr)
+
\frac{2\mu}{3}\Bigl(\frac{4\Delta^4}{\epsilon^5}-\frac{3\Delta^2}{\epsilon^3}\Bigr)\nonumber\\
U&=& \frac{4\xi}{9}\Bigl(\frac{-\Delta^4}{\epsilon^4}+\frac{\Delta^2}{\epsilon^2}\Bigr)
+\frac{4\mu}{9}\Bigl(\frac{-\Delta^4}{\epsilon^4}+\frac{\Delta^2}{\epsilon^2}\Bigr)\;.
\end{eqnarray}

For completeness we give the expression for $F$ which can be evaluated
similarly,
\begin{eqnarray}
F=\frac{1}{8\pi^2} \Biggl[
\int_{0}^{\magp_-} d\magp \frac{\magp^2\xi}{\epsilon^3} +
\int_{\magp_+}^{\infty} d\magp \frac{\magp^2\xi}{\epsilon^3} \Biggr]\label{F final}\;.
\end{eqnarray}

\section{TDGL equation for the vortex core states}
\label{ginsburg-landau}

To analyze the vortex core structure we derive the
Ginsburg-Landau functional  expanding the action
\begin{equation}
S = \int d^4x \frac{|\eta(x)|^2}{\lambda}-\frac{1}{2}{\rm Tr} \log\bf{S}^{-1} \,,
\end{equation}
 around a state with $\Delta=0$ up to the  fourth order in $\eta$, obtaining
\begin{equation}
\begin{split}
S_{eff}&=S^{(2)} + S^{(4)} = \frac{T}{V}\sum_k
\tilde{\eta}(-k)\tilde{\eta}^{*}(k)\frac{1}{\lambda}
+\left(\frac{T}{V}\right)^2\sum_k\sum_p
\tilde{\eta}(-k)\tilde{\eta}^{*}(k)
\frac{1}{4}\frac{2}{A(p)\tilde{A}(p+k)} \nonumber\\
&+\left(\frac{T}{V}\right)^2\sum_k\sum_p 
(\tilde{\eta}(-k)\tilde{\eta}^{*}(k))^2\frac{1}{8}
\frac{2}{(A(p)\tilde{A}(p+k))^2} + (\delta\mu\rightarrow -\delta\mu)\,,
\end{split}
\end{equation}
where $A,\tilde{A}$ are defined in Eq.~(\ref{define A}).
We rewrite the action as 
\begin{equation}
S^{(2)} + S^{(4)} = \frac{T}{V}\sum_k
\tilde{\eta}(-k)\tilde{\eta}^{*}(k) J_2(k)
+(\tilde{\eta}(-k)\tilde{\eta}^{*}(k))^2 J_4(k)\,,
\end{equation}
where $J_4$ can be written as 
\begin{eqnarray}
J_4(k) &=& -\frac{1}{4}\frac{T}{V}\sum_p
\frac{\partial}{\partial{\mu_1}}\frac{\partial}{\partial{\mu_2}}
\left(\frac{1}{A(p)\tilde{A}(p+k)}+(\delta\mu\rightarrow -\delta\mu)\right)\,.
\end{eqnarray}
We perform the $k_0$ Matsubara sum, take $p_0=0$, the limit ${\bf p}\rightarrow 0$ and  obtain the effective action 
\begin{equation}
S_{eff}=\int\frac{d^3 p}{(2\pi)^3}\left(\eta^2(a+c\frac{p^2}{2m}) + \eta^4 \frac{b}{2}\right) \,,
\end{equation}
where the coefficients are given by
\begin{eqnarray}
a &=& -\frac{m}{4\pi a_s} +\int \frac{d^3k}{(2\pi)^3}\Bigl[
\frac{1}{2\frac{k^2}{2m}}-\frac{1}{2\xi_k}(1-n_{+}-n_{-})
\Bigr]\,,\nonumber\\
c &=& \int\frac{d^3k}{(2\pi)^3}\Bigl[
\frac{1}{4\xi_k^2}(1-n_{+}-n_{-})+\frac{1}{4\xi_k}(n^{\prime}_{+}+n^{\prime}_{-})
\nonumber\\
&+&\frac{(k n)^2}{4m\xi_k}\Bigl(
-\frac{1}{\xi_k^2}(1-n_{+}-n_{-})
-\frac{1}{\xi_k}(n^{\prime}_{+}+n^{\prime}_{-})
+(n^{\prime\prime}_{+}+n^{\prime\prime}_{-})
\Bigr)
\Bigr]\,,\nonumber\\
b &=& \int\frac{d^3k}{(2\pi)^3}\Bigl[
\frac{1}{4\xi_k^3}(1-n_{+}-n_{-})
+\frac{1}{4\xi_k^2}(n^{\prime}_{+}+n^{\prime}_{-})
\Bigr]\,,
\label{coef}
\end{eqnarray}
where
$n_{\pm}=n(\xi_k\pm\delta\mu)$, $n$ is the Fermi distribution,
$n^{\prime}(x)=\frac{\partial n}{\partial x}$,
$\xi_k=\frac{k^2}{2m}-\mu$, $\mu=(\mu_1+\mu_2)/2$, $\delta\mu=(\mu_1-\mu_2)/2$
and we take $\mu_1>\mu_2$. 
The equation of motion, $\delta S_{eff}/\delta \eta(p) =0$, is given by
\begin{equation}
\left(a+b\,\eta(p)^2+c\frac{p^2}{2m}\right)\eta(p)=0\,,
\end{equation}
and in configuration space
\begin{equation}
\left(a+b\,\eta(r)^2-\frac{c}{2m}\nabla^2
\right)\eta(r)=0\,,
\end{equation}
with the boundary conditions $\eta(r=0)=0$, $\eta(r\rightarrow\infty)=\eta_0$.
At $T\neq 0$, we have
\begin{eqnarray}
a &=& -\frac{m}{4\pi a_s}+
\frac{(2m\Delta_0)^{3/2}}{\Delta_0}\frac{1}{16\pi^2}\Bigl(
2\int_0^{\infty}\frac{dx}{x^{1/2}}
-2\int_0^{\infty}\frac{x^{1/2}
dx}{x-\rho}[1-n(x-\rho_{+})-n(x-\rho_{-})]\Bigr)\nonumber\\
&\equiv& -\frac{(2m\Delta_0)^{3/2}}{\Delta_0}\frac{1}{16\pi^2}\,\tilde{a}
\nonumber\\
b &=& \frac{(2m\Delta_0)^{3/2}}{\Delta_0^3}\frac{1}{16\pi^2}\Bigl(
\int_0^{\infty}\frac{x^{1/2}dx}{(x-\rho)^3}[1-n(x-\rho_{+})-n(x-\rho_{-})]
\nonumber\\
&+&\int_0^{\infty}\frac{x^{1/2}dx}{(x-\rho)^2}[n^{\prime}(x-\rho_{+})+n^{\prime}(x-\rho_{-})]\Bigr)\nonumber\\
& \equiv & \frac{(2m\Delta_0)^{3/2}}{\Delta_0^3}\frac{1}{16\pi^2}\,\tilde{b}
\nonumber\\
c &=& \frac{(2m\Delta_0)^{3/2}}{\Delta_0^2}\frac{1}{16\pi^2}\Bigl(
\int_0^{\infty}\frac{x^{1/2}dx}{(x-\rho)^2}[1-n(x-\rho_{+})-n(x-\rho_{-})]
\nonumber\\
&+&\int_0^{\infty}\frac{x^{1/2}dx}{x-\rho}[n^{\prime}(x-\rho_{+})+n^{\prime}(x-\rho_{-})]\Bigr.\nonumber\\
&-&\Bigl.\frac{2}{3}\int_0^{\infty}\frac{x^{3/2}dx}{(x-\rho)^3}[1-n(x-\rho_{+})-n(x-\rho_{-})]
-\frac{2}{3}\int_0^{\infty}\frac{x^{3/2}dx}{(x-\rho)^2}[n^{\prime}(x-\rho_{+})+n^{\prime}(x-\rho_{-})]\Bigr.\nonumber\\
&+&\Bigl.\frac{2}{3}\int_0^{\infty}\frac{x^{3/2}dx}{x-\rho}[n^{\prime\prime}(x-\rho_{+})+n^{\prime\prime}(x-\rho_{-})]
\Bigr)
\equiv \frac{(2m\Delta_0)^{3/2}}{\Delta_0^2}\frac{1}{16\pi^2}\,\tilde{c}
\label{abc} 
\end{eqnarray}
where we introduced $T =\Delta_0$, and
the dimensionless variables $x=k^2/2m\Delta_0$, $\rho=\mu/\Delta_0$ and $\nu=\delta\mu/\Delta_0$;
zeros of the quasiparticle energy are at $\rho_{\pm}=\rho\pm\nu$.
   
Introducing $\eta(r)={\rm e}^{i\phi}f(\zeta)\eta_0$ with dimensionless $\zeta=r\sqrt{2m\eta_0}$,
we obtain the TDGL equation for the vortex core at $T\neq 0$ 
\be
 \tilde{c}\left(\frac{1}{\zeta}\frac{d}{d\zeta}\left(\zeta\frac{df}{d\zeta}\right)
-\frac{f}{\zeta^2}\right)+\tilde{a}f-\tilde{b}f^3=0\,, \label{tdgl1}
\ee
with boundary conditions $f(0)=0$ and $f(\infty)=1$. 
The coefficients $\tilde{a}$, $\tilde{b}$, $\tilde{c}$ are given in the right hand side of Eq.~(\ref{abc}).
Coefficients $\tilde{a}=\tilde{b}=\tilde{c}=1$ correspond to a superfluid ideal Bose gas discussed 
by Landau~\cite{landau}, where a vortex filament has macroscopic thickness. 
Here we are able to study both regimes of BCS and BEC, BCS-BEC transition, as well as the situation with
nonzero mismatch, $\delta\mu\neq 0$.  
We solve this second order ODE numerically for different values of $\delta\mu$.

\end{document}